\documentclass[pra,aps,showpacs,floatfix,twocolumn,superscriptaddress]{revtex4-2}
\usepackage{amsmath}
\usepackage{amsfonts}
\usepackage{amssymb}
\usepackage{revsymb}
\usepackage{graphicx}
\usepackage{physics}
\usepackage{mathtools}
\usepackage{tabularx}
\usepackage{booktabs}
\usepackage[table]{xcolor}
\usepackage{hyperref}

\graphicspath{{figs/}}

\begin{document}

\title{Analysis of collision shift assessments in ion-based clocks}
\author{M. D. Barrett}
\affiliation{Centre for Quantum Technologies, National University of Singapore, 3 Science Drive 2, 117543 Singapore}
\affiliation{Department of Physics, National University of Singapore, 2 Science Drive 3, 117551 Singapore}
\author{K. J. Arnold}
\affiliation{Centre for Quantum Technologies, National University of Singapore, 3 Science Drive 2, 117543 Singapore}
\affiliation{Temasek Laboratories, National University of Singapore, 5A Engineering Drive 1, Singapore 117411, Singapore}
\begin{abstract}
We consider back-ground gas collision shifts in ion-based clocks.  We give both a classical and quantum description of a collision between an ion and a polarizable particle with a simple hard-sphere repulsion.  Both descriptions give consistent results, which shows that a collision shift bound is determined by the classical Langevin collision rate reduced by a readily calculated factor describing the decoupling of the clock laser from the ion due to the recoil motion.  We also show that the result holds when using a more general Lennard-Jones potential to describe the interaction between the ion and its collision partner.  This leads to a simple bound for the collision shift applicable to any single ion clock without resorting to large-scale Monte-Carlo simulations or determination of molecular potential energy curves describing the collision.  It also provides a relatively straightforward means to measure the relevant collision rate.
\end{abstract}

\maketitle
\section{Introduction}
Optical atomic clocks based on single trapped ions have greatly improved in accuracy over the past two decades with several systems reporting inaccuracies at the $10^{-18}$ or beyond \cite{zhang2026liquid,brewer2019al+,hausser2025in+,marshall2025high,zhiqiang2023176lu+} and even demonstrating agreement to the same level \cite{zhiqiang2023176lu+}.  At the mid-$10^{-19}$ level, a potentially important systematic is the collision shift, which arises due to the collision of the clock ion with residual background molecules in the vacuum chamber.  Estimate of the shift is particularly important for Lu$^+$, for which uncertainties at the low $10^{-19}$ are now accessible \cite{zhiqiang2023176lu+} making the collision shift one of the most important and yet least well-understood systematic for this system.  

A variety of estimates for the collision shift have appeared in the literature with varying degrees of sophistication.  The most extensive accounts are a Monte-Carlo analysis given in \cite{hankin2019systematic}, which is primarily a classical treatment appropriate to collision energies of concern, or the quantum formulation given in \cite{vutha2017collisional,davis2019improved}.  At the opposite extreme, a bound on the collision shift can be made by ignoring all details of the collision dynamics and assume every collision has the worst possible impact on the spectroscopy signal.  For Ramsey spectroscopy, that would be to shift the phase by $\pm\pi/2$.  For a collision rate $\Gamma$, the Ramsey signal would then be
\[
\frac{1}{2}\left(1+e^{-\Gamma T}\cos(\Delta T)\pm(1-e^{-\Gamma T})\sin(\Delta T)\right)
\]
leading to a bound
\begin{equation}
\label{WCCS}
\frac{\delta f_0}{f_0}\approx \pm\frac{\Gamma}{2\pi f_0},
\end{equation}
where we have assumed $\Gamma T\ll1$, which is to assume at most one collision per interrogation cycle.  A similar result can be found for Rabi spectroscopy.  We will refer to this as the worst-case collision shift (WCCS).  It is a useful point of comparison when considering the importance of various factors contributing to a collision shift.

Classically, the collision rate between an ion and a polarizable particle is determined by Langevin scattering, which has a rate that is independent of the collision energy \cite{cote2000ultracold,zipkes2011kinetics}.  At vacuum levels of interest, the background gas is predominately molecular hydrogen for which the classical Langevin collision rate at $300\,\mathrm{K}$ is $\Gamma_L\approx 3.6 \times 10^{-4}\,\mathrm{s^{-1} nPa^{-1}}$ using the orientation averaged polarizability for molecular hydrogen given in \cite{hankin2019systematic}.  For Al$^{+}$, this gives a fractional collision shift bound of $\pm 5.1\times 10^{-20} \mathrm{nPa^{-1}}$.  This is within $\sqrt{2}$ of the bound given \cite{davis2019improved} as an improved estimate of the collision shift, which utilized a quantum formulation incorporating calculated potential energy curves for the AlH$_2^+$ complex.  Within a factor of 2 this is also the case for bounds given for Sr$^+$\cite{leibin2025collisional}, which were derived in a similar way.  Moreover, applying the result in  \cite{davis2019improved} to the 57\,nPa upper bound on the pressure used in \cite{hankin2019systematic} yields a bound of $\pm 2\times 10^{-18}$, which is about an order magnitude larger than that given in \cite{hankin2019systematic}.  If estimates of collision shifts are to be a meaningful part of error budgets, there should be reasonable consistency between different formulations.  Thus, it is important to investigate the differences and limitations of the two approaches, to understand where the differences arise, to determine what factors are important, and the extent that those factors may be specified.

In this paper we consider the two formulations in \cite{hankin2019systematic,davis2019improved} within the context of an idealized model of a small hard-sphere repulsion in addition to the long range attraction between an ion and a polarizable particle.  This allows analytic results to be derived, which provide insight into the importance of various factors included in the Monte-Carlo simulation reported in  \cite{hankin2019systematic}.  It is demonstrated that a collision shift bound is determined by the classical Langevin collision rate reduced by a readily calculated factor describing the decoupling of the clock laser from the ion due to its recoil motion, with all other factors being of secondary importance.  It is also shown this is consistent with the formulation given in \cite{davis2019improved} when taking into account the influence of the ion's recoil on the atom-laser coupling.  Consequently we are able to give a simple description of the collision shift that can be readily evaluated for any single-ion clock and eliminates the need for large scale Monte-Carlo simulations or accurate calculations of molecular potential energy curves.  We also provide insight into the level of accuracy needed in the calculation of molecular potential energy curves to improve upon the collision shift bound we give.

The paper is organized as follows.  In section I, we give a simple heuristic consideration of the collision which allows us to quantify the importance that collisions have on an atom-laser coupling.  This highlights the importance of the recoil velocity distribution resulting from the collision.  We quantify the distribution in section II, using a simplified classical treatment of Langevin scattering.  This indicates that ``glancing collisions'', which do not penetrate the angular momentum barrier and significantly increase the collision rate above the classical Langevin rate, have no significant role and can be ignored.  We justify this in Section III using a fully quantum mechanical treatment with an attractive $r^{-4}$ potential in conjunction with a hard-sphere repulsion.  This is further supported by replacing the hard-sphere repulsion with Lennard-Jones potentials, which reveals the collision shift sensitivity to the accuracy of calculated molecular potentials.  Our treatment considers Ramsey spectroscopy, but results could be readily adapted to Rabi spectroscopy.

\section{The Ramsey Suppression Factor}
\label{sectI}
We first start with a basic description of a collision between a background hydrogen molecule and an ion.  We focus on H$_2$ as it is the predominant background gas at typical vacuums used in ion trap experiments, but results can be readily adapted for other species.  Due to the energies involved, collisions are largely classical in nature and for the most part can be treated as such.  Under typical operating conditions, the collision rate will be sufficiently low that there will be at most one collision per interrogation cycle.  A single collision will result in a near instantaneous phase shift, which we will refer to as the collision phase, and a velocity kick to the ion, which can be subsequently described by a coherent state in each of three dimensions.  The induced motion of the ion will result in a further phase accumulation over the remainder of the Ramsey interrogation due to the second order Doppler shift (SODS), but, more importantly, it will diminish the coupling of the laser during the final Ramsey pulse.

When the ion is in a coherent state of motion, the atom-laser coupling is equivalent to using a modulated laser, which can be formally established starting with the Hamiltonian in the interaction picture
\begin{equation}
H_I=D\left(i \eta e^{-i \omega_T t}\right)\sigma^++\mathrm{c.c}
\end{equation}
where $D(\alpha)$ is the displacement operator, $\omega_T$ is the trap angular frequency, and $\eta$ the Lamb-Dicke parameter.  With an instantaneous velocity kick, a general state $\ket{\psi}$ becomes $D(\alpha)\ket{\psi}$ for $\alpha \in \mathbb{R}$.  The subsequent dynamics is best described using a frame transformation with the unitary operator $D(-\alpha)$ in which the original state $\ket{\psi}$ evolves under the Hamiltonian
\begin{multline}
D(-\alpha)H_I D(\alpha)\\
=D\left(i \eta e^{-i \omega_T t}\right)e^{2i\eta\alpha \cos(\omega_T t)}\sigma^++\mathrm{c.c},
\end{multline}
which is exactly that of a modulated laser.

For concreteness, we consider the clock laser to be propagating along the radial direction at 45 degrees to the principal axes of the trap, which is a common configuration employed in the lab. In this geometry, the carrier will be depleted by a factor
\begin{equation}
\label{eq:carrier}
\beta_0=J_0\left(\frac{k_L v \sin\theta_t\cos\phi_t}{\omega_x\sqrt{2}}\right)J_0\left(\frac{k_L v \sin\theta_t\sin\phi_t}{\omega_y\sqrt{2}}\right)
\end{equation}
where $J_0(x)$ is the zeroth order Bessel function of the first kind, $\theta_t$ and $\phi_t$ define the direction of the velocity kick relative to the trap axis, $\omega_x$ and $\omega_y$ are the trap frequencies in the radial direction, $k_L$ the laser wavenumber, and $v$ is the recoil velocity of the ion.  This assumes the trap frequencies are such that combinations of sidebands will not be near resonant with the transition i.e. there are no integers $n,m$ with $n \omega_x+m\omega_y\approx0$ to within the frequency resolution the clock laser $\pi$-pulse provides.  It is readily shown that this leads to a Ramsey signal given by
\begin{equation}
\label{Eq:CollisionRamsey}
P(\Delta)=\frac{1}{2}\left[1+\sin(\frac{\pi \beta_0}{2})\cos(\Delta T+\phi_1+\phi_2)\right],
\end{equation}
where $\phi_1$ is a phase shift from the collision dynamics and $\phi_2$ is the phase shift from the SODS.  The latter is given by
\[
\phi_2=\left(\frac{v}{2 c}\right)^2 \omega_0 (T-\tau)=\omega_2  (T-\tau),
\]
where $\tau$ is the time at which the collision occurs relative to the first Ramsey pulse, $\omega_0$ is the clock angular frequency, and $\omega_2$ is the angular frequency shift of the clock transition due to the SODS.  Averaging the cos term over $\tau$ with an exponential distribution $\Gamma e^{-\Gamma \tau}$ gives
\begin{multline}
\label{eq:timeaverage}
\int_0^T \cos(\Delta T+\phi_1+\phi_2)\Gamma e^{-\Gamma \tau}\,d\tau\\
=\mathcal{C}\cos(\Delta T+\phi_1)-\mathcal{S}\sin\left(\Delta T+\phi_1\right),
\end{multline}
where 
\begin{align*}
\mathcal{C}&=\sin\theta_2\cos\theta_2\sin(\omega_2 T)-\sin^2\theta_2(e^{-\Gamma T}-\cos(\omega_2 T))\\
\mathcal{S}&=\sin\theta_2\cos\theta_2(e^{-\Gamma T}-\cos(\omega_2 T))+\sin^2\theta_2\sin(\omega_2 T)
\end{align*}
and
\[
\cos(\theta_2)=\frac{\omega_2}{\sqrt{\omega_2^2+\Gamma^2}},\quad \sin(\theta_2)=\frac{\Gamma}{\sqrt{\omega_2^2+\Gamma^2}}.
\]
Strictly speaking the use of an exponential distribution for the distribution of collision times only applies when the collision rate is constant as is the case for Langevin scattering.  More generally, an energy dependent collision rate will lead to a compound distribution.  We assume that such a distribution can at least be reasonably approximated by an exponential, which can be shown to be the case for a hard sphere scattering for example.

The total Ramsey signal can then be written
\[
\tfrac{1}{2}\left[1+e^{-\Gamma T}\cos(\Delta T)+\mathcal{A}\sin(\tfrac{1}{2}\pi \beta_0)\cos(\Delta T+\phi_1+\eta_v)\right],
\]
where $\mathcal{A}=\sqrt{\mathcal{C}^2+\mathcal{S}^2}\leq1-e^{-\Gamma T}$, and $\tan \eta_v=\mathcal{S}/\mathcal{C}$.  Note that for small velocities $\mathcal{A}\approx \mathcal{C}\approx 1-e^{-\Gamma T}$ reflecting the fact that the SODS is insufficient to diminish the fringe contrast.

The angles $\phi_t$ and $\theta_t$ define a direction relative to a fixed laboratory frame.  As the incoming direction is isotropic so is the recoil direction, independent of the differential scattering cross-section.  Consequently they are uncorrelated with $\phi_1$ and $\eta_v$.  We can then average over all $\theta_t$ and $\phi_t$, which replaces $\sin(\tfrac{1}{2}\pi \beta_0)$ with a factor $\mathcal{R}$, which we will refer to as the Ramsey suppression factor (RSF).

Averaging over velocities, $v$, would require knowledge of $\phi_1$ and the distribution of $v$, which depends on details of the collision dynamics.  For simplicity, we may bound the shift by taking $\phi_1+\eta_v=\pm\pi/2$, which would maximize the shift.   Hence, a bound on the clock shift including velocity effects would be given by
\begin{equation}
\label{eq:shiftFull}
\left|\frac{\delta f_0}{f_0}\right|<\frac{\langle |\mathcal{A}\mathcal{R}|\rangle_v}{2\pi f_0 T},
\end{equation}
where the average is over the ion recoil velocity distribution and we take the absolute value as we are bounding the maximum possible shift.

For collisions with a background gas of neutral atoms with mass $m_n$ at a temperature $T_0$, the energy transfer to the ion can be quantified by a dimensionless parameter $\delta$ given by
\begin{equation}
\delta=\left(\frac{v}{\tilde{v}}\right)^2=\frac{1}{2}\left(\frac{v_n}{\tilde{v}_n}\right)^2 \sin^2\left(\tfrac{\theta}{2}\right)
\end{equation}
where $v_n$ is the velocity of the neutral having a scale $\tilde{v}_n=\sqrt{k_B T_0/m_n}$, $v$ is the final velocity of the ion with mass $M$ having a velocity scale $\tilde{v}=\sqrt{2\beta k_B T_0/M}$, and $\beta$ given by
\begin{equation}
\label{eq:MaxTransfer}
\beta=\frac{4 M m_n}{(M+m_n)^2}
\end{equation}
is the maximum energy transfer via a head on collision.  Thus $\delta$ is the energy of the ion in units of $k_B$ relative to the temperature $\beta T_0$.  For convenience, we will henceforth use non-dimensional velocities $\bar{v}=v/\tilde{v}$ for the ion recoil velocity and $\bar{v}_n=v_n/\tilde{v}_n$ for the incoming velocity of the neutral.  Strictly, $\theta$ is the scattering angle in the center-of-mass frame.  As we are primarily interested in $^{176}$Lu$^+$ colliding with H$_2$, this distinction is of little consequence.  Moreover, it can be accounted for as maybe necessary when considering very light ions or heavy vacuum contaminants.

\begin{figure}[t]
\begin{center}
\includegraphics[width=0.95\linewidth]{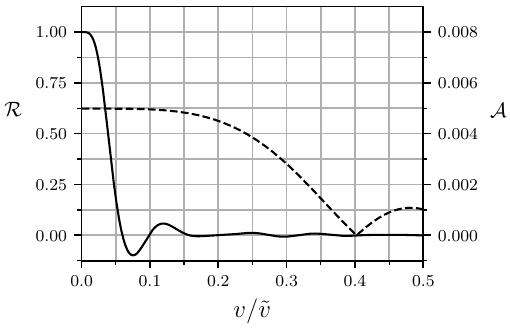}
\caption{\label{fig:Ramsey} Plots of the factors $\mathcal{R}$ (solid) and $\mathcal{A}$ (dashed) given in the text.  We have taken a background gas of molecular hydrogen, a collision rate of $\Gamma=1/1000\,\mathrm{s}^{-1}$, an interrogation time $T=5\,\mathrm{s}$ and $\omega_x=\omega_y=2\pi\times 500\,\mathrm{kHz}$.}
\end{center}
\end{figure}

For a background gas of hydrogen molecules at a temperature of $T_0=300\,\mathrm{K}$ colliding with $^{176}\mathrm{Lu}^+$, $\beta T_0=13.3\,\mathrm{K}$, and $\tilde{v}\approx 35.5\,\mathrm{ms^{-1}}$ and we will use these values as default unless otherwise stated.  In Fig.~\ref{fig:Ramsey} we plot $\mathcal{R}$ and $\mathcal{A}$ as a function of $\bar{v}$, where we have taken a collision rate of $\Gamma=1/1000\,\mathrm{s}^{-1}$, an interrogation time $T=5\,\mathrm{s}$, $\omega_x=\omega_y=2\pi\times 500\,\mathrm{kHz}$ and $k_L$ the wavenumber for the clock laser with wavelength $\lambda=848\,\mathrm{nm}$.  Note that $\omega_x \neq \omega_y$ is required for Eq.~\ref{eq:carrier} to hold, but the average value is not sensitive to the difference.  

As is evident from Fig.~\ref{fig:Ramsey}, $\mathcal{A}$ is practically flat over the region in which $\mathcal{R}$ is appreciable, and maybe replaced by $(1-e^{-\Gamma T})\approx \Gamma T$ in this limit.  This reflects the fact that the SODS is not sufficient to influence the overall averaging. In this limit, the clock shift becomes
\begin{equation}
\label{eq:RSCS}
\left|\frac{\delta f_0}{f_0}\right|\approx\frac{\Gamma}{2\pi f_0}\langle |\mathcal{R}|\rangle_v,
\end{equation} 
which is just the WCCS suppressed by the velocity-averaged RSF.  This approximation would suffice for most purposes, but would become less valid at longer interrogation times as $\mathcal{A}$ would have an increasingly larger influence at smaller velocities.  But that effect would be to decrease the estimate so Eq.~\ref{eq:RSCS} remains a valid bound.

The Ramsey suppression factor $\mathcal{R}$ has the physical interpretation that the final Ramsey pulse is completely ineffective or decoupled from the ion for sufficiently large recoil velocities, and hence the collision removes the interrogation from the Ramsey signal.  This interpretation must be taken in the context of an average over multiple collisions having different directions of recoil with respect to the clock laser propagation direction.  Any given collision will have some small contribution to the Ramsey signal in accordance with Eq.~\ref{Eq:CollisionRamsey} but these contributions tend to cancel when averaged over recoil directions and more so for larger recoil velocities.  This is due to the fact that the modulation indices appearing in Eq.~\ref{eq:carrier} are typically large.  So not only is $\beta_0$ typically small, but it also quickly averages to zero over recoil orientations.

If we make the simplifying assumption that the recoil velocity is uncorrelated with the phase $\phi_1$, and certainly that was implicitly assumed in the Monte-Carlo simulations presented in \cite{hankin2019systematic}, we can further average the Ramsey signal over the velocities.  For the purposes of estimation, we can take $\mathcal{R}$ to be a step function with a cutoff at $\bar{v}_{c}\ll1$.  Over this scale the probability density function (PDF) for the ion-recoil velocity can be expanded to lowest order which is taken to be either $\kappa \bar{v}$ or simply $\kappa$ where $\kappa$ is a number on the order of unity.  That is to say that the PDF is assumed to be either linear in the neighbourhood of zero or approximately constant.  For these two cases, the collision part of the Ramsey fringe can be shown to be well approximated by
\begin{subequations}
\label{eq:RSCSapprox}
\begin{equation}
\label{eq:RSCSapproxA}
\tfrac{1}{2}\Gamma T \kappa \bar{v}_{c}^2 \cos(\Delta T+\phi_1 +\tfrac{1}{2}\gamma \bar{v}_{c}^2)
\end{equation}
if the PDF is linear in the neighbourhood of zero, or
\begin{equation}
\label{eq:RSCSapproxB}
\Gamma T \kappa \bar{v}_c \cos(\Delta T+\phi_1 +\tfrac{1}{\pi}\gamma \bar{v}_{c}^2)
\end{equation}
if it approximately constant.  In both cases 
\begin{equation}
\gamma=\left(\frac{\tilde{v}}{2c}\right)^2\omega_0 T.
\end{equation}
\end{subequations}

Averaging over a recoil velocity distribution will always result in an expression given in Eqs.~\ref{eq:RSCSapprox} for an appropriately chosen $\bar{v}_c$ and scale factor for the SODS phase $\gamma \bar{v}_c^2$ under the assumed forms of the recoil velocity PDF.  In \cite{hankin2019systematic}, the authors take a worst case collision phase $\phi_1=\pm\pi/2$ citing \cite{rosenband2008frequency}, in which it is shown that a fixed value of $\pi/2$ gives the worst case result. But motional effects were not included in \cite{rosenband2008frequency}.  If the SODS was to lead to an effective shift of $\pi/2$ in Eqs.~\ref{eq:RSCSapprox}, then the choice of $\phi_1=\pm\pi/2$ would no longer be the worst case but rather the most optimistic.  Since the phase from the SODS appearing in Eqs.~\ref{eq:RSCSapprox} is typically $\ll \pi/2$, the choice of $\pm\pi/2$ would lead to basically the same bound with a small bias that one might then attribute to the SODS.  Consequently, the numbers given in \cite[Table I]{hankin2019systematic}, which are ascribed to phase and SODS contributions are really just an artefact of the choice of phase made.  The worst case choice consistent with  \cite{rosenband2008frequency} would be to compensate the small contribution from the SODS, rendering it inconsequential beyond its contribution to $\mathcal{A}$, which would be negligible in most realistic interrogation scenarios. 

The worst-case approach to the collision shift by maximizing the shift of the Ramsey fringe is likely extreme.  From a purely classical perspective, a collision phase shift would come from energy level differences between the clock states integrated over the collision trajectory.  For the typical scales involved in molecular potentials and room temperature collisions, one would anticipate collision phases being much larger than $2\pi$ and broadly distributed.  Any continuous distribution of phases that is broad with respect to $2\pi$, will give an approximately uniform distribution when taken modulo $2\pi$.  Thus the collision shift would tend to average to zero.  To see this one can take the phase $\phi_1$ as a normally distributed random variable with mean $\phi_0$ and standard deviation $\sigma$.  Averaging over $\phi_1$ then replaces $\phi_1$ with the mean $\phi_0$ and further suppresses the contribution to the Ramsey signal by a factor $e^{-\sigma^2/2}$.  It's hard to imagine $\sigma$ would be small.

Evidently, the key consideration for the collision shift is suppression of the Ramsey signal and consequently the recoil velocity distribution.  To give a concrete example as a point of reference, we consider hard-sphere collisions for which the differential cross section is constant and independent of the incoming velocity.  Using the Maxwell-Boltzmann (MB) speed distribution
\begin{equation}
\label{eq:MBSpeed}
P(\bar{v}_n)=\sqrt{\frac{2}{\pi}}\bar{v}_n^2 e^{-\tfrac{\bar{v}_n^2}{2}}
\end{equation} 
the probability $g(\delta)$ for a resulting recoil energy of up to $\delta k_B \beta T_0$ is given by
\begin{equation}
g(\delta)=\frac{1}{\sqrt{2\pi}}\int_0^\pi \int_0^{\bar{v}_{n,\delta}} \sin\theta \bar{v}_n^2 e^{-\tfrac{\bar{v}_n^2}{2}} d\bar{v}_n d\theta
\end{equation}
where 
\begin{equation}
\label{eq:ucut}
\bar{v}_{n,\delta}= \frac{\sqrt{2\delta}}{\sin\left(\tfrac{\theta}{2}\right)}.
\end{equation}
The probability distribution function (PDF) for $\delta$ is then obtained by differentiating with respect to $\delta$ giving
\begin{equation}
P(\delta)=2\,\mathrm{erfc}\sqrt{\delta},
\end{equation}
where $\mathrm{erfc}(x)$ is the complementary error function.  Using $P(\delta)d\delta=P(\bar{v})d\bar{v}$, this may be re-expressed as a velocity distribution, which is given by
\begin{equation}
\label{eq:HSV}
P(\bar{v})=4\bar{v}\,\mathrm{erfc}\left(\bar{v}\right).
\end{equation}
\begin{figure}[t]
\begin{center}
\includegraphics[width=0.95\linewidth]{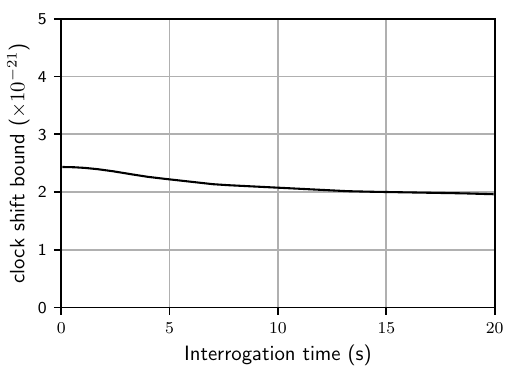}
\caption{\label{fig:clkshift} Collision shift bound for different interrogation times using a hard-sphere collision model.  Other parameters are as for Fig.~\ref{fig:Ramsey}.}
\end{center}
\end{figure}
In Fig.~\ref{fig:clkshift}, we plot the collision shift bound as a function of interrogation time calculated via Eq.~\ref{eq:shiftFull}.  The slight decrease with interrogation time is due to variations in $\mathcal{A}$.  The value is consistent with that determined from Eq.~\ref{eq:RSCSapproxA} using $\kappa=4$ derived from Eq.~\ref{eq:HSV} and $\bar{v}_c\approx 0.034$.  The value of $\bar{v}_c$ is slightly less than the point at which $\mathcal{R}\approx0.5$, which occurs at $\bar{v}_c \approx 0.039$ for the geometry we have considered.

We note in passing that a recoil velocity distribution for the ion is given in \cite{hausser2025in+} but no indication is given as to what the distribution is or how it was derived.  As the simple hard-sphere example shows, the distribution depends on the scattering properties and will not inherit the MB form of the background collision partner.  To do so, would require scattering at a fixed angle.  Moreover, the functional dependence at small recoil energies or velocities is qualitatively changed.  This is further highlighted in the next section, which shows that the velocity distribution resulting from a polarizable particle colliding with an ion is roughly a half gaussian making Eq.~\ref{eq:RSCSapproxB} the more relevant average.

\section{Classical Langevin Scattering}
\label{sect:Classical}
The classical description of a polarizable particle colliding with an ion was first described by Langevin.  The interpretation given in the literature is that the attractive $r^{-4}$ potential results in a separation of collisions: those above a critical impact parameter result in glancing collisions with minimal deflection, and those below the critical impact parameter result in an inward-spiralling trajectory that results in a hard collision between the two particles.  It is instructive to solve the problem classically to illustrate the behaviour.

\begin{figure}[t]
\begin{center}
\includegraphics[width=0.9\linewidth]{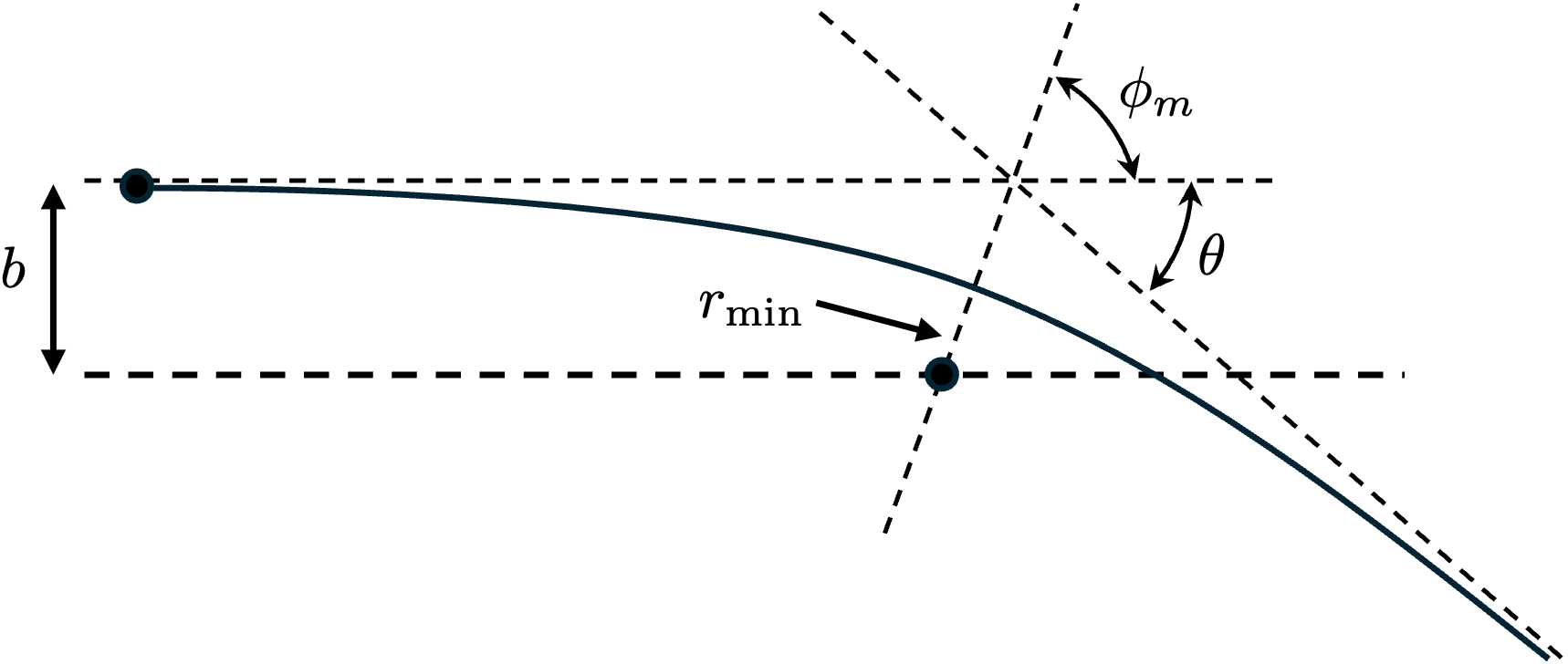}
\caption{\label{fig:collision} Typical collision for an attractive potential showing the impact parameter $b$, distance of closest approach $r_\mathrm{min}$, and angles $\phi_m$ and $\theta$ given in the text.}
\end{center}
\end{figure}

For any central potential $V(r)$, the orbit has the formal solution \cite{fetter2003theoretical}
\begin{equation}
\label{Eq:orbit}
\phi=\phi_0\pm\sqrt{\frac{l^2}{2m}}\int\frac{dr}{r^2\sqrt{E-V(r)-\tfrac{l^2}{2m r^2}}}
\end{equation}
where $l$ and $E$ are, respectfully, the angular momentum and energy, which are constants for elastic collisions.  For a polarizable particle colliding with an ion, the potential is given by
\begin{equation}
V(r)=-\frac{C_4}{2r^4}
\end{equation}
where $C_4=q^2 \alpha/(4\pi \epsilon_0)^2$, $q$ is the ion charge, and $\alpha$ the polarizability of the neutral particle.  Note that the factor of 2 in the expression for the potential is sometimes absorbed into the definition of $C_4$ and should be accounted for in expressions for the critical impact parameter e.g.\cite{harter2014cold} and \cite{zipkes2011kinetics}.

Substituting the potential into Eq.~\ref{Eq:orbit}, defining the impacter parameter, $b$, and velocity, $v_\infty$, by
\[
l=mv_\infty b,\quad E=\tfrac{1}{2} m v_\infty^2,\\
\] and scaling $r$ by the critical impact parameter $b_c= (2 C_4/E)^{1/4}$ leads to the non-dimensional form
\begin{align}
\phi&=\phi_0\mp\int \frac{2\bar{b}du}{\sqrt{\left(u^2-2\bar{b}^2\right)^2+4(1-\bar{b}^4)}},\\
&=\phi_0\mp\mathcal{I}(u,\bar{b})
\end{align}
where $u=1/\bar{r}$ has been substituted and $\bar{b}=b/b_c$.

When $b>b_c$, $u$ is restricted to the region $0<u<u_m$ where 
\[
u_m=\bar{b}\sqrt{2(1-\sqrt{1-\bar{b}^{-4}})}
\]
or, equivalently, $\bar{r}>\bar{r}_m$ where 
\[
\bar{r}_m=u_m^{-1}=\bar{b}\sqrt{(1+\sqrt{1-\bar{b}^{-4}})/2},
\]
and a typical trajectory is shown in Fig.~\ref{fig:collision}, which identifies parameters of interest.  

For algebraic purposes, it is convenient to set $\bar{b}^2=\cosh x$ for $x>0$, which gives $u_m^2=2e^{-x}$, and the integral is given by
\begin{multline}
\label{eq:orbit}
\mathcal{I}(u,\bar{b})=\sqrt{1+e^{-2x}}\, F\left(\sin^{-1}\left(\tfrac{u e^{x/2}}{\sqrt{2}}\right),e^{-2x}\right)\\
=\sqrt{1+\tfrac{1}{4}u_m^4}\, F\left(\sin^{-1}\left(\tfrac{u}{u_m}\right),\tfrac{1}{4}u_m^4\right)
\end{multline}
where $F(\Phi,m)$ is the incomplete elliptic integral of the first kind defined by
\[
F(\Phi,m)=\int_0^\Phi\frac{d\theta}{\sqrt{1-m \sin^2\theta}}.
\]

Assuming the incoming particle has an angle $\phi=\pi$, the angle $\phi_m$ at the minimum approach is given by 
\[
\phi_m=\pi-\sqrt{1+\tfrac{1}{4}u_m^4}\, K(\tfrac{1}{4}u_m^4)
\] 
and the scattering angle $\theta=\pi-2 \phi_m$ where $K(m)=F(\pi/2,m)$ is the complete elliptic integral of the first kind.  Thus, $\theta$ diverges to $\infty$ as $b\rightarrow b_c$ meaning the incoming particle can spiral around the ion before scattering back outwards.  However, this only occurs for impact parameters very close to $b_c$ e.g. for $b=1.00225 b_c$, $\theta\approx \pi$ which corresponds to a back reflection of the neutral.

When $b<b_c$, $u$ (or $\bar{r}$) is no longer restricted and the particle will ``spiral'' into the ion.  But, as for $b>b_c$, the spiralling behaviour only occurs for impact parameters very close to $b_c$.  Setting $\bar{b}=\sqrt{\cos(x)}$ for $x\in(0,\pi/2)$ gives $u_m^2=2e^{-i x}$ and Eq.~\ref{eq:orbit} still holds with $x$ replaced with $i x$.  The resulting expression is, of course, still real.  This can be shown by using an ascending Landen transformation followed by a reciprocal modulus transformation, which gives
\begin{equation}
\label{Eq:orbit2c}
\mathcal{I}(u,\bar{b})=\sqrt{2\cos x}\,F\left(\Phi,\cos^2\left(x/2\right)\right),
\end{equation}
where
\[
\sin(\Phi)=\sec(x/2)\sin\left(\Re\left(\sin^{-1} Z\right)\right),
\]
and $Z=u e^{i x/2}/\sqrt{2}$. This leads to the limiting value of
\begin{equation}
\sqrt{2\cos x}\,K\left(\cos^2\tfrac{x}{2}\right)=\bar{b}\sqrt{2}\,K\left(\tfrac{1+\bar{b}^2}{2}\right)
\end{equation}
for $u\rightarrow \infty$ i.e. $\bar{r}\rightarrow 0$.

\begin{figure*}[t]
\begin{center}
\includegraphics[width=0.32\linewidth]{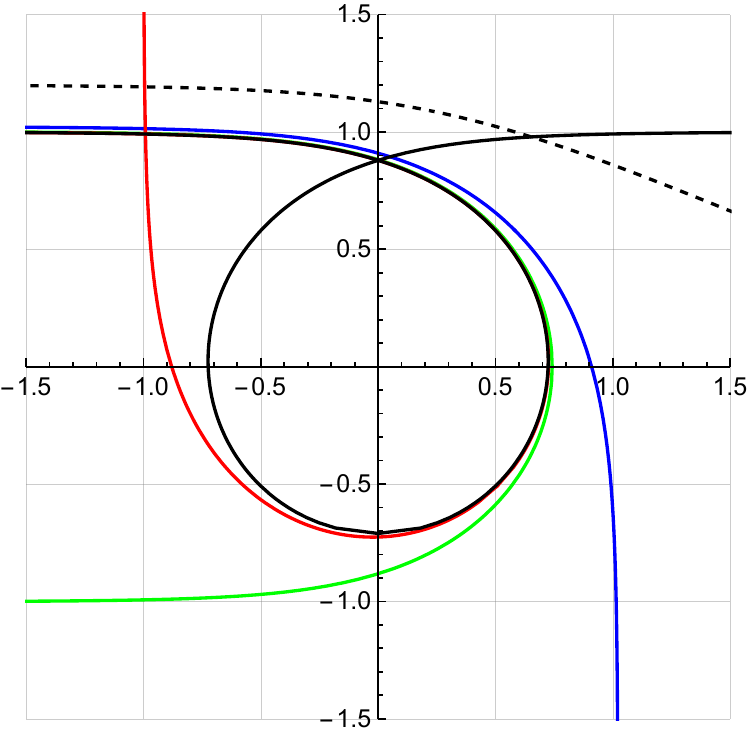}
\includegraphics[width=0.32\linewidth]{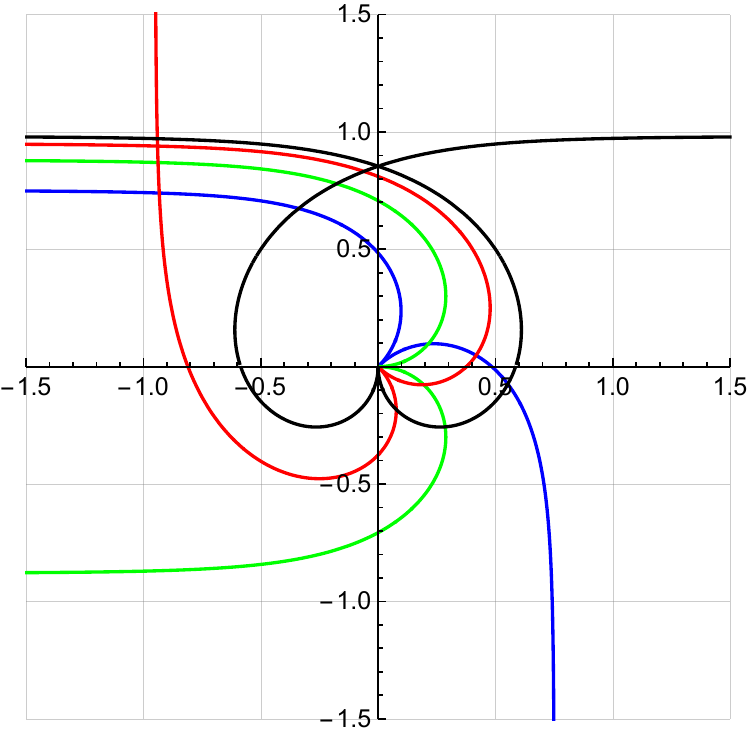}
\includegraphics[width=0.32\linewidth]{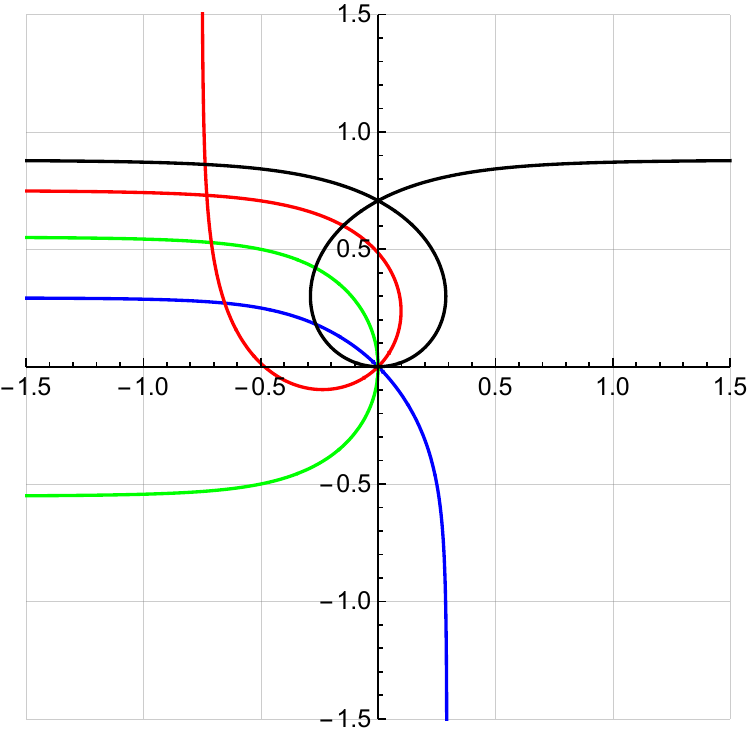}
\caption{\label{fig:orbits} Orbits for scattering angles $\theta=\pi/2,\pi,3\pi/2,$ and $2\pi$ for different scenarios with a fixed energy.  In all cases axes are scaled by the critical parameter $b_c$: (left) $b>b_c$.  The dashed curve has $b=1.2b_c$; (middle) $b<b_c$ where it is assumed there is a sharp repulsion that reflects the particle at small $r$; (right) $b<b_c$ where it is assume the trajectory passes through the origin.}
\end{center}
\end{figure*}

\begin{figure}[t]
\begin{center}
\includegraphics[width=0.9\linewidth]{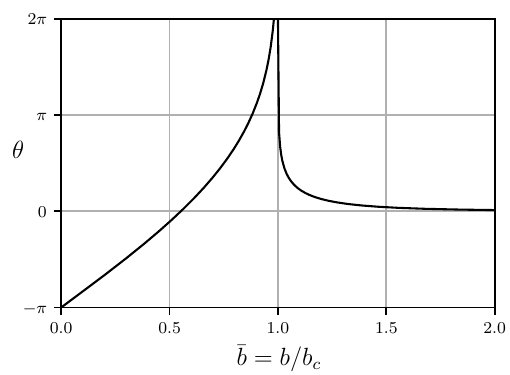}
\caption{\label{fig:scattering} Scattering angle as a function of impact parameter where the impact parameter is scaled by $b_c$.  For $b<b_c$ the scattering angle assumes a sharp repulsion at small $r$.}
\end{center}
\end{figure}

In Fig.~\ref{fig:orbits}, we illustrate the orbits that arise for different impact factors for a fixed energy.  Impact parameters for both $b>b_c$ and $b<b_c$ were chosen to give deflections of $\pi/2,\pi,3\pi/2,$ and $2\pi$ to illustrate the onset of spiralling behaviour.  For $b<b_c$, the orbit passes through $r=0$, which requires the velocity to go to infinity to conserve angular momentum.  We take a simplest model of the collision in which it is assumed there is a sharp repulsive potential or hard-sphere that directly reflects the neutral at $r\approx 0$, but, for completeness, we plot trajectories for both cases. More realistic behaviour would need to be modelled by the appropriate intermolecular potentials, but the simple reflection is amenable to analytic calculations. For $b>b_c$, the case of $b=1.2b_c$ is also given to illustrate how the deflection falls off with increasing $b$.  It is readily seen that orbits that spiral about the ion only occur for $b\approx b_c$, which is further illustrated in Fig.~\ref{fig:scattering} where we plot the scattering angle as a function of $b/b_c$. Scattering angles above $2\pi$ indicate a full orbit, which only falls in the narrow range of $b\in (0.981,1.000)b_c$ assuming there is a sharp repulsion at small $r$.

We note in passing that any $l\neq0$ supports a closed orbit.  When $b>b_c$, the closed orbit is a circle of radius $\bar{r}_m$ centered on the origin.  Thus the corresponding open orbit intersects tangentially with the closed orbit at the point of minimum approach.  When $b<b_c$, the closed orbits are circular orbits that pass through the origin with radius $R=b_c^2/(4b)$ i.e. $r=2 R \cos(\phi-\phi_0)$.  In this case the open orbit intersects tangentially with the closed orbit at the origin.  

\subsection{The collision rate}
For Langevin scattering, the total cross section is taken as $\sigma=\pi b_c^2$, which neglects glancing collisions for which $b>b_c$.   Consequently, the total collision rate $n\sigma v_n$ is independent of the collision velocity $v_n$.  As in \cite{hankin2019systematic}, it is of interest to consider a differential rate for collisions that impart an energy $E_i=\delta k_B \beta T_0$ to the ion.  For convenience, we will express this in terms of the non-dimensional $\delta$.

The total collision rate imparting up to $E_i=\delta k_B \beta T_0$ to the ion is given by
\begin{align}
\Gamma&=2\pi \sqrt{\tfrac{2}{\pi}}n \tilde{v}_n  \int_0^\pi \int_0^{\bar{v}_{n,\delta}} \bar{v}_n^3 e^{-\bar{v}_n^2/2} \sin\theta \frac{d\sigma}{d\Omega}d\bar{v}_nd\theta\\
&=2\pi \sqrt{\tfrac{2}{\pi}}n \tilde{v}_n \int_0^\pi \int_0^{\bar{v}_{n,\delta}} \bar{v}_n^3 e^{-\bar{v}_n^2/2} b \left|\frac{d b}{d\theta}\right|d\bar{v}_nd\theta
\end{align}
where $n$ is the background gas density and $\bar{v}_{n,\delta}$ is given by Eq.~\ref{eq:ucut}.  It is convenient to introduce an impact parameter scale, $\tilde{b}_c$, which is found by using $\tilde{v}_n$ in the expression for $b_c$.  This gives $b_c^2=\tilde{b}^2_c/\bar{v}_n$ and the collision rate is then
\begin{equation}
\Gamma=\Gamma_L\sqrt{\tfrac{8}{\pi}}\int_0^\pi \int_0^{\bar{v}_{n,\delta}} \bar{v}_n^2 e^{-\bar{v}_n^2/2} \left(\bar{b} \left|\frac{d \bar{b}}{d\theta}\right|\right)d\bar{v}_nd\theta,
\end{equation}
where $\Gamma_L=n \pi \tilde{b}_c^2 \tilde{v}_n$ is the classical Langevin collision rate.  Differentiating with respect to $\delta$ then gives the desired differential collision rate
\begin{multline}
\label{eq:dBGC}
\frac{d\bar{\Gamma}}{d\delta}=\sqrt{\tfrac{16\delta}{\pi}}\int_0^\pi e^{-\delta \csc^2(\theta/2)}\csc^3(\theta/2) \left(\bar{b} \left|\tfrac{d \bar{b}}{d\theta}\right|\right)d\theta,
\end{multline}
where $\bar{\Gamma}=\Gamma/\Gamma_L$.

In this expression, the scattering angle shown in Fig.~\ref{fig:scattering}, should be rescaled to the interval $(0,\pi)$, which replaces the bracketed expression in the integral with a summation to account for the fact that multiple impact parameters give effectively the same scattering angle.  For $b<b_c$, it can be shown that the bracketed term, as a function of $\theta\in(0,\pi)$, is well approximated to $p_0+p_1\cos(\theta/2)$ with the parameter regions over which multiple orbiting occurs having a diminishing contribution.  Substituting $p_0+p_1\cos(\theta/2)$ and integrating over $\theta$ then gives
\begin{equation}
\label{eq:LE}
\frac{d\bar{\Gamma}}{d\delta}=\left[4p_0M\left(\delta\right)+\frac{4p_1}{\sqrt{\pi\delta}}e^{-\delta}\right].
\end{equation}
The function $M(x)=G(\{\{\},\{0\}\},\{\{-1/2,1/2\},\{\}\},x)$ is a Meijer G function, and the term in square parentheses is a normalized distribution function with an integrable singularity at $\delta=0$ scaling as $\delta^{-1/2}$.  We can also express this in terms of scaled velocity $\bar{v}$, which amounts to multiplying by $2\sqrt{\delta}$ and substituting $\delta=\bar{v}^2$ for which we get
\begin{equation}
\label{eq:LV}
\frac{d\bar{\Gamma}}{d\bar{v}}=\left[8 p_0 \bar{v} M\left(\bar{v}^2\right)+\frac{8 p_1}{\sqrt{\pi}}e^{-\bar{v}^2}\right].
\end{equation}
The exact values of $p_0$ and $p_1$ are inconsequential to the purposes of discussion but they must be consistent for normalization.  We take $p_0=1/(3\pi)$ and $p_1=1/12$, which are close to the fitted values, and gives a weighting of (2/3) and (1/3) to the two terms.  Note that the fitted values give a numerical normalization of $1$, which serves as a check on the derivation.

\begin{figure}[t]
\begin{center}
\includegraphics[width=0.9\linewidth]{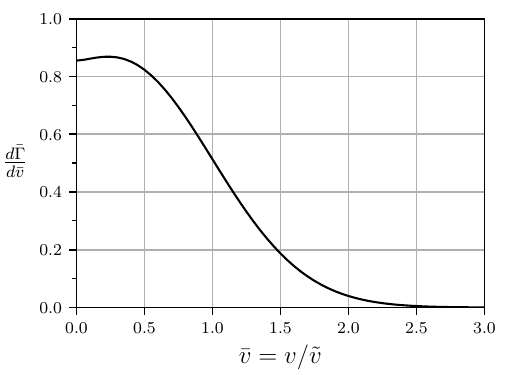}
\caption{\label{fig:LV} Differential background collision rate (Eq.~\ref{eq:LV}) as a function of scaled velocity $(\bar{v})$.}
\end{center}
\end{figure}

The above considerations have neglected glancing collisions.  Inasmuch as this is valid,  the collision rate is independent of the incoming energy and the probability distribution functions for recoil energy or velocity will be as given in Eq.~\ref{eq:LE} and Eq.~\ref{eq:LV} respectively.  Consequently, the only change to an estimated clock shift given in the previous section is in the PDF for the recoil velocities.  As shown in Fig.~\ref{fig:LV}, the velocity distribution is non-zero for $\bar{v}\rightarrow 0$, which would tend to increase the influence on a Ramsey fringe.

Glancing collisions ($b>b_c$) should have a reduced effect on clock accuracy.  Such collisions do not penetrate the angular momentum barrier and, consequently, any phase shift would have limited dependence on the clock state i.e. $\phi_1=0$ in Eq.~\ref{eq:timeaverage}.  This fact was also incorporated into the Monte-Carlo simulations reported in \cite{hankin2019systematic}.  The predominant effect would then be the SODS and a change in the associated collision rate.  We note that this is distinctly not the case for neutral atoms, in which long range behaviour is governed by the state dependent $r^{-6}$ term.  As a consequence, the collision shift in neutral atom clocks is heavily dependent on forward scattering to which glancing collisions heavily contribute \cite{gibble2013scattering}.

Classically the total collision rate for glancing collisions diverges since the potential does not strictly vanish outside some finite distance.  Although this can be avoided by using a full quantum treatment, the scattering angle for $b>2 b_c$ is well approximated by $3\pi/(16\bar{b}^{4})$, and hence the influence of such collisions very quickly diminishes.  It is therefore reasonable to consider a cutoff in the ion's recoil energy below which collisions have negligible effect.

Although the contribution to the total collision rate from $b>b_c$ diverges, the contribution to the differential collision rate does not.   The required integral is still given by Eq.~\ref{eq:dBGC}, but with different expressions for $\bar{b}$.  As $\theta$ is given as a function of $\bar{b}=\sqrt{\cosh x}$, the integral can be converted to an integral over $x$ for which a good approximation can be found by using $e^{-x/2}$ as an expansion parameter for the $\csc(\theta/2)$ terms and keep only the leading order terms.  Specifically, we use 
\[
\csc(\theta/2)\approx -\frac{8}{3\pi}e^{2x},\quad \bar{b}\frac{d\bar{b}}{dx}=\frac{1}{2}\sinh x
\]
which yields
\begin{equation}
\label{eq:GE}
\frac{d\bar{\Gamma}}{d\delta}=\frac{128\sqrt{\delta}}{27\pi^{7/2}}\left[E_{-\frac{3}{4}}\left(\frac{64\delta}{9\pi^2}\right)-E_{-\frac{1}{4}}\left(\frac{64\delta}{9\pi^2}\right)\right],
\end{equation}
where $E_n(z)$ is the generalized exponential integral function.  The expression given was found by integrating over the full interval $x\in(0,\infty)$ and thus includes the small contribution from orbiting. It does not have the correct asymptotic behaviour for large $\delta$, but there is negligible contribution for $\delta>1$ and this is of no further consequence.  More importantly, the approximation has a non-integrable singularity at $\delta=0$ with a power law $\sim \delta^{-5/4}$, and is hence not normalizable.  It maybe normalized by introducing a cutoff for $\delta$. 

The integral
\begin{equation}
\label{eq:classical_cutoff}
N_c=\int_{\delta_c}^\infty \frac{d\bar{\Gamma}}{d\delta}\,d\delta.
\end{equation}
represents the relative contribution to the total collision rate from glancing collisions resulting in recoil energies with $\delta>\delta_c$ and is a normalization factor for a truncated distribution from Eq.~\ref{eq:GE}.  Recall that if we specify the ion recoil energy as $E_i=k_B T_i$, then $\delta=T_i/(\beta T_0)=(v/\tilde{v})^2$, where $\beta T_0=13.3\,\mathrm{K}$ and $\tilde{v}=35.5\,\mathrm{ms^{-1}}$ for $^{176}$Lu$^+$ in a background gas of molecular hydrogen at 300\,K.  For recoil energies below $k_B\times1$\,mK the SODS is fractionally $<2.5\times 10^{-19}$ of the clock transition frequency and there is minimal effect on the laser coupling.  For these energies, the contribution is better considered as a collision-induced heating, which is already accounted for in error budgets through measured heating rates to which they would contribute.  Such collision-induced heating has been discussed in the literature for neutral atom traps \cite{bali1999quantum}.  Thus, the predominant contributions to the Ramsey signal comes from energies $T_i>1\,\mathrm{mK}$ for which $N_c\approx 5$.

\subsection{The collision shift revisited}
To include glancing collisions with energies above $\sim 1\,\mathrm{mK}$, we can use the same treatment as in Sect.~\ref{sectI}, with a velocity distribution determined from Eq.~\ref{eq:GE} truncated at $\sim1\,\mathrm{mK}$, and omitting the phase factor $\phi_1$ in Eq.\ref{eq:timeaverage}.  The resulting Ramsey signal can then be averaged with that derived for Langevin scattering with each weighted by probabilities determined from their corresponding contribution to the overall scattering rate.  The component arising from glancing collisions is given by
\[
(1-p)\left[\langle \mathcal{C R}\rangle_G\cos(\Delta T)-\langle \mathcal{S R}\rangle_G\sin(\Delta T)\right]
\]
where  $p=1/(1+N_c)$, and $\langle X \mathcal{R}\rangle_G$ are averages over the normalized velocity distribution for glancing collisions.  As the SODS has only a minor role in the averaging relative to the influence motion has on the laser coupling, the second term is negligible, which is equivalent to treating $\mathcal{A}$ as a constant.  Consequently, the glancing component plays no significant role except for a negligible loss of the fringe contrast.  The Ramsey signal is then given by basically the same expression as in the previous section.  Specifically,
\begin{equation}
\tfrac{1}{2}\Big(1+e^{-\Gamma T}\cos(\Delta T)\pm p \langle \left|\mathcal{A R}\right|\rangle_L\sin(\Delta T)\Big)
\end{equation}
where $\Gamma=\Gamma_L(1+N_c)$, and $\langle \left|\mathcal{A} \mathcal{R}\right|\rangle_L$ is the average over the normalized velocity distribution for Langevin (L) collisions.  As before we take the absolute value as we consider the worst case phase shift from the collision and the $\pm$ gives the two possible extremes.  As in the previous section $\mathcal{A}$ can be taken to be practically constant and given by $\Gamma T$.  The increased rate is of no consequence as $p\Gamma=\Gamma_L$. The fractional clock shift is then bounded by Eq.~\ref{eq:RSCS} as before with the rate $\Gamma$ still determined by $\Gamma_L$.  That is to say the glancing collisions have no significant contribution.  Any small bias arising from the neglected sin terms, as seen in \cite{hankin2019systematic}, would always be compensated by the correct choice of worst case collision phase $\phi_1$ with little change to the estimated bound.

For a background gas of molecular hydrogen at a pressure of 3\,nPa with temperature $T_0=300\,\mathrm{K}$, $\Gamma_L\approx 0.001\,\mathrm{s}$.  With a cutoff energy in the range 0.3-2\,mK, and an interrogation time of 5\,s, we find $\pm1.4\times 10^{-20}$ for the collision shift, which is insensitive to the choice of cutoff and insensitive to the interrogation time.  The increase in the estimate relative to that given in the previous section is due to the change in the velocity distribution, which is now non-zero at $v=0$.  The value is consistant with $\kappa\approx 0.855$ from Eq.~\ref{eq:LV}  and $\bar{v}_c=0.034$ as before.

We note that differentiating the probability distribution between glancing and Langevin collisions is a purely classical argument.  The distributions are also differentiated in \cite{hankin2019systematic} as they use two different differential cross-sections, found with and without glancing collisions, in determining the probabilities assigned to the two different collisions implemented in the Monte-Carlo simulation.  Independent of how the differential cross-sections are constructed, when separated in this fashion the result is distinctly classical.  Quantum mechanics does not allow this separation as the addition of the two contributing scattering amplitudes results in interference effects \cite{bransden2003physics}.  We explore the consequences of this in Sect.~\ref{sect:Quantum}.

\subsection{Collison-induced heating}
The truncated part of the distribution in Eq.~\ref{eq:GE} has no significant effect on the Ramsey signal, or on any heating, which would be given by
\begin{equation}
\Gamma_H=\left(\Gamma_L \beta T_0\right) \int_0^{\delta_c} \frac{d\bar{\Gamma}}{d\delta} \delta\,d\delta.
\end{equation}
A cutoff at low energies is not needed as the singularity is now integrable.  However, we have introduced a cutoff at high energies to illustrate the effect that high energy transfers have and the negligible contribution from the divergence at low energies, which was previously omitted.  High energy transfers dominate the heating rate in spite of their infrequent occurrence.  How this should be considered in reality is somewhat a matter of opinion and we include the result as a matter of completeness to illustrate the negligible role of the diverging collision rate from glancing collisions.  The scale factor $\Gamma_L \beta T_0$ is independent of the background temperature and dependent on the ion species, background gas, and pressure.  For $^{176}$Lu$^+$ in a background gas of molecular hydrogen, it is $4.77\,\mathrm{mK\, s^{-1} nPa^{-1}}$.

\begin{figure}[t]
\begin{center}
\includegraphics[width=0.95\linewidth]{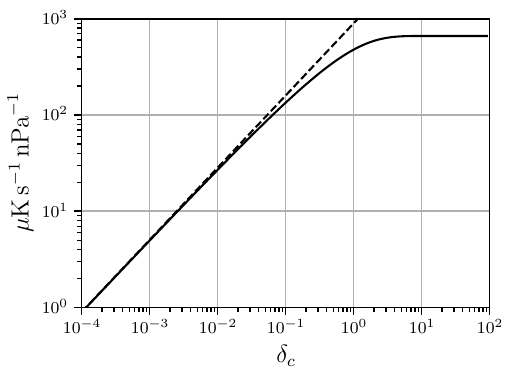}
\caption{\label{fig:heating} Heating rate as a function of the cutoff $\delta_c$ calculated for Lu$^+$ in a background gas of molecular hydrogen at 300\,K.  The dashed line gives the approximate power law, which is valid for small angle/low energy deflections.}
\end{center}
\end{figure}

The heating rate per nPa is plotted in Fig.~\ref{fig:heating} as a function of $\delta_c$ assuming a background gas of molecular hydrogen at 300\,K.  For $\delta_c\ll 1$ the heating rate is well approximated by a power law $895.5 \delta_c^{3/4}\,\mathrm{uK\,s^{-1}nPa^{-1}}$ under the assumed conditions.  Note that this expression gives a background temperature dependence since $\delta_c$ is expressed as a fraction of $\beta k_B T_0$.  The heating rate saturates as $\delta_c$ increases as high energy transfers have an increasingly small cross-section and are suppressed by the MB distribution.  For the cutoff of 1\,mK used in the previous section with a background pressure of $3\,\mathrm{nPa}$, the heating rate is just $2.2\,\mathrm{\mu K\,s^{-1}}$, which illustrates the negligible contribution of the divergence to a heating rate.

\subsection{Collision-induced Stark-shift}
We can carry out a similar analysis to the previous section for determining a phase shift induced on a superposition of two clock states.  We consider the Stark shift from the induced dipole on the incoming molecule.  The total shift will be determined by an integral over time, however, as the field from the induced dipole falls off rapidly with distance and energies are typically high, we can take the integral over the entire orbit with the assumption that the vast majority of the shift happens in a small time window when the molecule is close to the atom.  It is implicitly assumed the timescale is long enough that the dynamics of the electronic structure can be neglected, i.e. the a Born approximation applies, and that we can assume the shift is described by the static polarizability.  In addition, we assume the tensor component of the polarizability averages to zero.

The electric field on the ion from the induced dipole will be given by
\[
\mathbf{E}=\frac{2p}{4\pi \epsilon r^3}\hat{\mathbf{r}},
\]
where the induced dipole $p$ is given by
\[
p=\alpha \frac{q}{4\pi \epsilon_0 r^2}.
\]
Defining
\[
\mathcal{E}=\frac{2 \alpha q}{(4\pi\epsilon_0)^2},
\]
the phase shift on the clock superposition is given by
\begin{align}
\Delta\phi&=-\frac{\Delta\alpha_0 \mathcal{E}^2}{2\hbar} \int \frac{1}{r^{10}} \frac{dt}{dr}\,dr\nonumber\\
&=\frac{\Delta\alpha_0 \mathcal{E}^2}{\hbar b_c^{10}}\frac{1}{\bar{b}v_n}\int_0^{u_\mathrm{m}} \frac{u^8 du}{\sqrt{\tfrac{1}{\bar{b}^2}-u^2+\tfrac{1}{4\bar{b}^2}u^4}}\\
&=\frac{\Delta\alpha_0 \mathcal{E}^2}{\hbar b_c^{10}}\frac{1}{v_n} F(x).
\end{align}
As indicated, the integral is best carried out using $\bar{b}=\sqrt{\cosh x}$.  The rate at which phase is accumulated is then
\begin{align}
\frac{d\Delta\phi}{dt}=&\int_0^\infty \int_0^\pi \Delta \phi n v_n b \frac{d b}{d \theta} P(v_n) d\theta dv_n\nonumber\\
=&\int_0^\infty \int_0^\infty \Delta \phi n v_n b_c^2 \, \bar{b} \frac{d \bar{b}}{d x} P(v_n) dx dv_n\nonumber\\
=&\frac{n \Delta\alpha_0 \mathcal{E}^2}{\hbar \tilde{b}_c^7}\sqrt{\tfrac{2}{\pi}}\int_0^\infty u^{11/2} e^{-u^2/2}\,du\nonumber \\
&\quad\times\int_0^\infty \tfrac{1}{2}F(x)\sinh(x)\,dx,\nonumber\\
=&\frac{n \Delta\alpha_0 \mathcal{E}^2}{\hbar \tilde{b}_c^7}\kappa,
\end{align}
where $\kappa\approx 3.5$.
For a 1\,nPa background pressure of hydrogen at 300\,K, we get $7.1\times 10^{-11} \,\mathrm{s}^{-1}$.  Since this is the accumulation per unit time, it is equated directly with a detuning and hence a clock shift of
\[
\delta f=\frac{1}{2\pi}\frac{d\Delta\phi}{dt},
\]
or $\sim 3.2 \times 10^{-26}$ fractionally.  In \cite{hankin2019systematic}, this effect was stipulated to be negligible due to the size of $\Delta\alpha_0$.  However, we see that it remains negligible under rather extreme increases in both background vacuum pressure and $\Delta\alpha_0$.  Consequently, any such phase shift has to come from differences in the intermolecular potentials \cite{vutha2017collisional}, which requires impact parameters $b\lesssim b_c$. 

\section{Quantum Treatment}
\label{sect:Quantum}
The classical treatment of glancing collisions in the previous section is ad hoc.  Based on considerations of the uncertainty principle, they require a quantum description \cite{berman1982collision}.  To this end, we consider a quantum description of the $-r^{-4}$ potential with a hard-sphere repulsion.  A complete quantum solution for the $-r^{-4}$ potential is given in \cite{o1961modification} specifically for the purpose of matching solutions to short range scattering potentials to those of the long range $-r^{-4}$ potential.  A similar construction is given in \cite{idziaszek2011multichannel}, which also provides an algorithm for numerically calculating the required scattering phase.  Here we provide a summary of the relevant properties and show that a modified Wentzel–Kramers–Brillouin (WKB) solution, similar to that given in \cite{dalgarno1958mobilities}, provides an excellent analytic approximation to quantum solution, at least for the collision energies of interest here.  We then show that the collision shift derived in \cite{vutha2017collisional,davis2019improved} does not depend on the contribution from glancing collisions and, when suitably modified to account for the atom laser decoupling, gives a consistent result with the classical result of the previous section.
\subsection{Solutions and approximations}
For the $-r^{-4}$ potential, the radial equation takes the form
\[
-\frac{d^2 u}{d r^2}+\left(\frac{l(l+1)}{r^2}-\frac{L^2}{r^4}\right)u=k^2 u,
\] 
where
\[
k=\sqrt{\frac{2m_nE}{\hbar^2}},\quad \mbox{and} \quad L=\sqrt{\frac{m_n C_4}{\hbar^2}}.
\]
The parameters $k$, $L$, and $b_c=(2 C_4/E)^{1/4}$ are related through a single parameter $q=\sqrt{2m_n^2 E C_4/\hbar^4}$ for which we have 
\[
 q=2\left(L/b_c\right)^2=kL=\tfrac{1}{2}\left(kb_c\right)^2.
 \]
When $r$ is scaled by $b_c/\sqrt{2}$ the radial equation has the non-dimensional form
\[
-\frac{d^2 u}{d \bar{r}^2}+\left(\frac{l(l+1)}{\bar{r}^2}-\frac{q}{\bar{r}^4}\right)u=q u,
\]
which has a critical value of $q_0=l(l+1)/2$. For $q>q_0$ ($q<q_0$), the corresponding energy is above (below) the angular momentum barrier.  For an energy of $k_B\times300\,\mathrm{K}$, $q_0\approx 370$ which corresponds to $l\approx 27$.

Substituting $u=w\sqrt{\bar{r}}$ and $\bar{r}=e^z$, transforms the radial equation to the modified Mathieu equation
\begin{equation}
\label{eq:ModMathieu}
\frac{d^2 w}{d z^2}-\left[a-2 q \cosh(2z)\right]w=0, \quad a=\left(l+\tfrac{1}{2}\right)^2.
\end{equation}
Solutions to this equation have been extensively studied over the last century.  A thorough discussion of mathematical properties is given in \cite{meixner1954spharoidfunktionen} and a summary of key results can be found on the NIST Digital Library of Mathematical Functions (\href{https://dlmf.nist.gov/28}{DLMF\S 28}).  For completeness, we give a summary of key properties relevant to our needs.

For a given pair $a, q$ there is a set of characteristic exponents $\nu=\pm\hat{\nu}+2n$, $n\in\mathbb{Z}$ where $0\leq \Re(\hat{\nu})\leq 1$ in which the solution may be expressed in the form
\[
w(z)=\sum_{m=-\infty}^\infty c^\nu_{2m}(q) e^{-(\nu+2m)z}.
\]
The two signs in the expression for $\nu$ reflect the fact that if $w(z)$ is a solution so is $w(-z)$, and the offset by $2n$ is a mere relabelling of the coefficients ($m\rightarrow m+n$).  Substitution into the differential equation generates a recursion relation between the $c^\nu_{2m}(q)$ that may be expressed in the form of an infinite tridiagonal matrix equation.  For a non-trivial solution, the determinant of the matrix must be zero, which defines an equation for $\nu$ that may be solved recursively \cite{strang2005characteristic}.

Of interest are solutions denoted $M_\nu^{(j)}(z,h)$ given in \href{https://dlmf.nist.gov/28.20}{DLMF\S 28.20} where $h=\sqrt{q}$, $j=1,2,3,4$ and
\begin{align*}
M_\nu^{(3)}(z,h)&=M_\nu^{(1)}(z,h)+i M_\nu^{(2)}(z,h),\\
M_\nu^{(4)}(z,h)&=M_\nu^{(1)}(z,h)-i M_\nu^{(2)}(z,h),
\end{align*}
For $z\rightarrow +\infty$, the asymptotic behaviour of these two solutions is given by
\[
M_\nu^{(j)}(z,h)\sim \frac{\exp\left[\pm i\left(2 h \cosh z-\left(\tfrac{1}{2}\nu+\tfrac{1}{4}\right)\pi\right)\right]}{\sqrt{2 \pi h}\cosh(z/2)},
\]
where $+(-)$ corresponds to $j=3$ ($j=4$).  In terms of the radial wavefunctions, these have the desired behaviours of incoming and outgoing waves.  To determine the behaviour as  $z\rightarrow -\infty$ ($\bar{r}\rightarrow 0$), it is convenient to express the two solutions in terms of $M_{\pm\nu}^{(1)}(z,h)$, which are linearly independent when $\nu$ is non-integer.  From \href{https://dlmf.nist.gov/28.22}{DLMF\S 28.22} 
\begin{multline}
M_\nu^{(1)}(z,h)\pm i M_\nu^{(2)}(z,h)\\
= \frac{\pm i\left[e^{\mp i \nu \pi} M_\nu^{(1)}(z,h)-M_{-\nu}^{(1)}(z,h)\right]}{\sin(\nu \pi)}
\end{multline}

As $M_\nu^{(1)}(z,h)$ is the sum of $M_\nu^{(3)}(z,h)$ and $M_\nu^{(4)}(z,h)$, the asymptotic behaviour for $z\rightarrow +\infty$ is given by
\begin{equation}
\label{asymptote1}
M_\nu^{(1)}(z,h)\sim \sqrt{\frac{2}{\pi h}}\frac{\cos\left(2 h \cosh z-\left(\tfrac{1}{2}\nu+\tfrac{1}{4}\right)\pi\right)}{\cosh(z/2)}.
\end{equation}

The asymptotic form for $M_\nu^{(1)}(z,h)$ along the negative $z$ axis can then be determined from the connection formula \href{https://dlmf.nist.gov/28}{DLMF\S 28}
\begin{equation}
M^{(1)}_{\nu}(-z,h)=\frac{M^{(1)}_{\nu}(0,h)}{M^{(1)}_{-\nu}(0,h)}M^{(1)}_{-\nu}(z,h),
\end{equation}
where we have used \href{https://dlmf.nist.gov/{28.22.13}}{DLMF\S {28.22.15}},
\href{https://dlmf.nist.gov/28.20.5}{DLMF\S {28.20.5}}, and \href{https://dlmf.nist.gov/28.12.8}{DLMF\S {28.12.8}}.  

\begin{widetext}
Two solutions to Eq.~\ref{eq:ModMathieu} are given in \cite{o1961modification}, which in our notation and scaling are
\begin{subequations}
\label{eq:ModSolution}
\begin{align}
u_{ps}(\bar{r})&=\sqrt{\frac{\pi \bar{r}}{2h\cos2\delta}}\left[m_{-\nu} \cos\delta \,M_{-\nu}^{(1)}\left(\ln(\bar{r})\right)+(-1)^l m_\nu \sin\delta\,M_{\nu}^{(1)}\left(\ln(\bar{r})\right)\right]\\
u_{pc}(\bar{r})&=\sqrt{\frac{\pi \bar{r}}{2h\cos2\delta}}\left[m_{-\nu} \sin\delta \,M_{-\nu}^{(1)}\left(\ln(\bar{r})\right)+(-1)^l m_\nu \cos\delta\,M_{\nu}^{(1)}\left(\ln(\bar{r})\right)\right],
\end{align}
where $m_{\nu}=M^{(1)}_{-\nu}(0,h)/M^{(1)}_{\nu}(0,h)$ and $\delta=\tfrac{1}{2}\pi\left(\nu-l-\tfrac{1}{2}\right)$.  For $\bar{r}\ll 1$, they have the asymptotic forms
\begin{equation}
u_{ps}(\bar{r})\sim\tfrac{\bar{r}}{h}\sin\left(\tfrac{h}{\bar{r}}-\tfrac{1}{2}l\pi\right), \quad \mbox{and}\quad u_{pc}(\bar{r})\sim\tfrac{\bar{r}}{h}\cos\left(\tfrac{h}{\bar{r}}-\tfrac{1}{2}l\pi\right).
\end{equation}
\end{subequations}
\end{widetext}
For $\bar{r}\gg1$, $u(r)=u_{ps}(\bar{r})+B u_{pc}(\bar{r})\propto \sin(h\bar{r}-\tfrac{1}{2}l \pi+\eta_l) $ where
\[
\tan(\eta_l)=\frac{m_{-\nu}^2-\tan^2\delta+B \tan\delta (m_{-\nu}^2-1)}{\tan\delta(1-m_{-\nu}^2)+B(1-m_{-\nu}^2\tan^2\delta)}.
\]
Thus, specifying $B$ sets the scattering phase.  With 
\[
B=-\tan\left(\tfrac{L}{R}-\tfrac{1}{2}\pi l\right),
\]
and $\bar{r}\ll1$,
\[
u(\bar{r})\sim \tfrac{\bar{r}}{h}\sin(\tfrac{h}{\bar{r}}-\tfrac{L}{R}),
\]
which is zero for $r=R$.  Hence, for sufficiently small $R$ so that the asymptotic form holds, a given $B$ specifies a hard-sphere repulsion at least over the energies of interest.  Moreover,  $L/R$ need only be specified modulo $\pi$ in this limit.

With $B$ specified it remains to determine $\nu$ and $m_\nu$.  As shown in \cite{strang2005characteristic}, the characteristic exponent satisfies
\begin{equation}
\label{eq:characteristic}
\sin^2\left(\tfrac{\pi \nu}{2}\right)=\Delta(a,q)\sin^2\left(\tfrac{\pi \sqrt{a}}{2}\right)=\frac{1}{2}\Delta(a,q).
\end{equation}
where $\Delta(a,q)$ is the determinant of an infinite tridiagonal matrix for which a recursion relation is given in \cite{strang2005characteristic} for computing it.  To avoid possible confusion, $\Delta$ as given  in \cite{strang2005characteristic}, is a function of $a,q,$ and $\nu$, but defines an equation for $\nu$, which can be expressed in terms of the value of $\Delta$ at $\nu=0$ denoted $\Delta(a,\nu=0)$ in \cite{strang2005characteristic}.  Here we indicate the dependence on the pair $a,q$ with the understanding that $\Delta(a,q)$ is the determinant at $\nu=0$ in Eq.~\ref{eq:characteristic}.  The value of $m_\nu$ requires the coefficients $c_{2m}^\nu$, which can be found using the algorithm in \cite[Appendix A]{idziaszek2011multichannel}.  The treatment in \cite{idziaszek2011multichannel} is equivalent to that in \cite{o1961modification} with only minor differences in the solutions.  The value of $m_\nu$ in Eqs.~\ref{eq:ModSolution} is the same as that given in \cite{idziaszek2011multichannel}.  When solving for $c_{2m}^\nu$ for a given $\nu$, it is better to choose an offset $2n$ to maximise $|c_0^\nu|$.

For the energies of interest, phase shifts can be adequately determined by the WKB approximation for which we have
\begin{multline}
\eta_l=\left(l+\tfrac{1}{2}\right)\tfrac{\pi}{2}-k r_0\\+\int_{r_0}^\infty\left(\left(k^2-\frac{2m_nV(r)}{\hbar^2}-\frac{\left(l+\tfrac{1}{2}\right)^2}{r^2}\right)^{1/2}-k\right)dr
\end{multline} 
where $r_0$ is the classical turning point \cite{bransden2003physics} and the Langer modification $l(l+1)\rightarrow(l+\tfrac{1}{2})^2$ has been used \cite{langer1937connection}.  Scaling $r$ by $b_c$ and using $u=1/\bar{r}$ gives 
\begin{multline}
\eta_l=\left(l+\tfrac{1}{2}\right)\tfrac{\pi}{2}-k b_c/u_0\\
+k b_c\int_{0}^{u_0}\left(\left(1-u^2 p^2+\tfrac{1}{4}u^4\right)^{1/2}-1\right)\frac{1}{u^2}du
\end{multline} 
where $u_0=b_c/r_0$ and $p=(l+\tfrac{1}{2})/(k b_c)$.  Analytic expressions for a hard-sphere of fixed radius was given in \cite{dalgarno1958mobilities}.   Although their integral appears different, it can be shown to be equivalent \cite{bransden2003physics}.  Integrations can be done in an analogous manner to the orbit integrals in the previous section.

When $p>1$, the classical turning point corresponds to $u_m$ as given in Sect.~\ref{sect:Classical} with $p$ replacing $\bar{b}$, and 
\begin{multline}
\label{WKBbelow}
\eta_l=\left(l+\tfrac{1}{2}\right)\tfrac{1}{2}\pi\\+\frac{k b_c}{u_m}\left[K\left(\tfrac{1}{4}u_m^4\right)\left(1-\tfrac{1}{4}u_m^4\right)-2E\left(\tfrac{1}{4}u_m^4\right)\right],
\end{multline}
which agrees with that in \cite{dalgarno1958mobilities}.  Strictly, this expression requires $R<r_0$, but as we are considering a limit of small $R$ we take this as given. Using
\[
\left(l+\tfrac{1}{2}\right)\tfrac{1}{2}\pi=\frac{\pi}{2}k b_c p=\frac{\pi}{2}\frac{k b_c}{u_m}\sqrt{1+\tfrac{1}{4}u_m^4},
\]
it can be shown that $\eta_l\approx \tfrac{\pi}{32}k b_c/p^3$ for $p\gg 1$ ($u_m\rightarrow 0$), which is precisely that given by \cite[Eq. 15]{cote2000ultracold}, and limits to $(\pi-\sqrt{8})k b_c/2$ as $p\rightarrow 1$.  

When $p<1$, the energy is above the angular momentum barrier and the phase shift becomes
\begin{multline}
\label{WKBabove1}
\eta_l=\tfrac{1}{2}\pi k b_c p-\tfrac{1}{4}\pi\\
+k b_c\sqrt{8}\left[F\left(\Phi,x\right)\left(1-x\right)-E\left(\Phi,x\right)\right],\\
+kb_c \left(u_0-\frac{\sqrt{4-4p^2u_0^2+u_0^4}}{2u_0}\right)
\end{multline}
where $x=(1+p^2)/2$, $u_0=b_c/R$, and
\[
\sin\Phi=\frac{\sqrt{u_0^2+2u_0\sqrt{2 x}+2}-\sqrt{u_0^2-2u_0\sqrt{2 x}+2}}{2\sqrt{2x}}.
\]
One can apply duplication formulae for the elliptic integrals \href{https://dlmf.nist.gov/19.11}{DLMF\S {19.11}} to the result in \cite{dalgarno1958mobilities}.  This gives a similar result to the above with the $-\pi/4$ removed and the third line changed to
\[
kb_c \left(u_0-\frac{u_0\sqrt{4-4p^2u_0^2+u_0^4}}{2+u_0^2}\right).
\]
The subtraction of $\pi/4$ arises from the hard wall, which changes the connection formulae in the WKB approximation.
The term on the third line determined from \cite{dalgarno1958mobilities} asymptotes to a finite value as $u_0$ increases, whereas our expression increases as $k b_c u_0/2=k b_c^2/(2R)=L/R$.  Since the integrand will asymptote to $\sim 1/r^2$ as $r\rightarrow 0$, it must diverge as $u_0$ increases.  In addition our expression agrees with numerical evaluation of the integral.  Hence we conclude that there is an error in \cite{dalgarno1958mobilities}.

Since we have defined the boundary condition on $B$ asymptotically for small $R$, the WKB solution should be treated in the same way.  For any given $R=R_0$, there is a sequence of $R_n<R_0$ limiting to zero such that $L/R_n=L/R_0+n\pi$ where $n\in \mathbb{Z}^+$ and hence give equivalent solutions.  The limit of this sequence on Eq.~\ref{WKBabove1} has $\Phi\rightarrow \pi/2$, giving complete elliptic integrals on the second line, and the terms on the third line replaced by the limit $L/R_0+n\pi = k b_c u_0/2+n\pi$.  Since the integer multiple of $\pi$ has no relevance, the limiting phase shift is then
\begin{multline}
\label{WKBabove2}
\eta_l=\tfrac{1}{2}\pi k b_c p-\tfrac{1}{4}\pi+L/R_0\\
+k b_c\sqrt{8}\left[K\left(x\right)\left(1-x\right)-E\left(x\right)\right],
\end{multline}
Thus, within this approximation, the only dependence on the size of the core is in the constant $L/R_0$, which can be specified modulo $\pi$ without loss of generality.   It is henceforth to be understood that $L/R$ is specified in this limit and denote the phase by $\phi_R$.  This phase is effectively the same as the constraint used by Case to keep the Hamiltonian for the $r^{-4}$ potential self adjoint \cite{case1950singular}.  As noted by Case, the constraint on the phase was equivalent to specifying a hard sphere center to avoid the singularity.

\begin{figure*}[t]
\begin{center}
\includegraphics{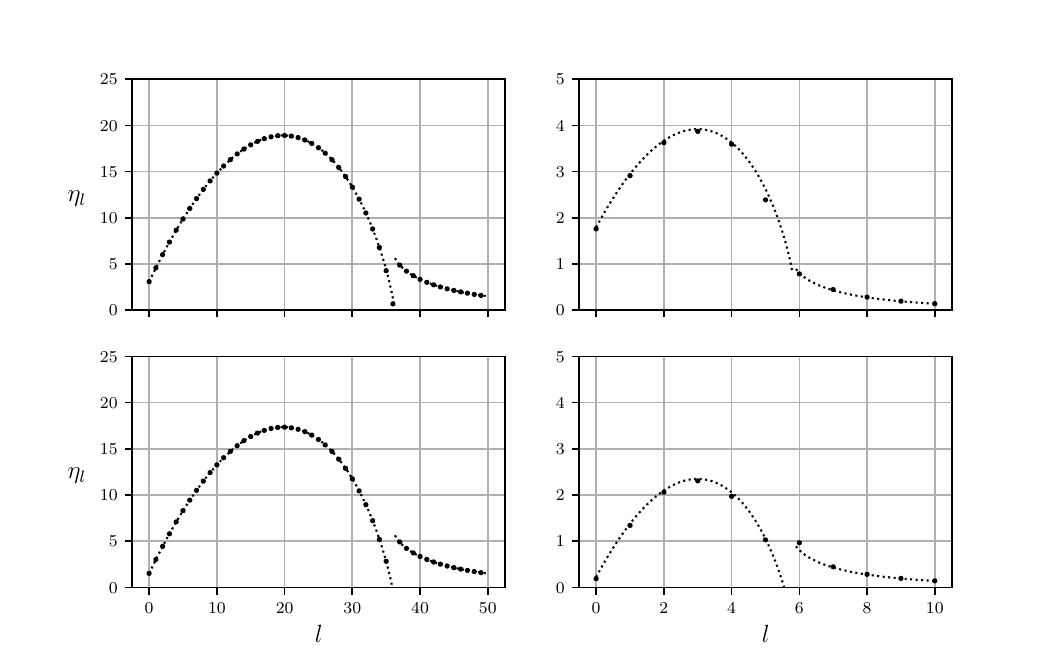}
\caption{\label{fig:WKB} Phase shifts calculated with the full quantum solution (dots) and the modified WKB approximation (dotted lines).  Left (right) plots have $q=675$ ($q=20$) corresponding to a collision energy $E\approx k_B\times 1000\,\mathrm{K}$ ($E\approx k_B\times 1\,\mathrm{K}$). Upper (lower) plots have $\phi_R=0$ $(\phi_R=\pi/2)$.  Points and WKB curves are offset by integer multiples of $\pi$ to illustrate the match.}
\end{center}
\end{figure*}

Equation~\ref{WKBabove2} gives an excellent match to the quantum solution over a wide range of energies and for all $p<1$ with the possible exception of one or two values of $l$ giving the largest $p<1$.  This is illustrated in Fig.~\ref{fig:WKB} in which we plot $\eta_l$ from the full quantum solution and the WKB solution for two collision energies and two different core radii.  Right and left plots have collision energies of $k_B\times 1\,\mathrm{K}$ and $k_B\times 1000\,\mathrm{K}$ respectively.  Upper and lower plots have $\phi_R= 0$ and $\pi/2$ respectively.

The matching of the WKB approximation to the quantum solution and the simple dependence it has on the core radius is a consequence of the energies involved.  It can be readily established that this is not the case in general, but becomes approximately true when tunnelling and above barrier reflection is negligible. This is typically the case except for very low energies not considered here or when the energy happens to be very near to the critical value.  The most significant deviations are for $l$ near to the critical value $l_0$ but the deviation is barely discernible on the plots.  

\subsection{Scattering properties}
\begin{figure}[t]
\begin{center}
\includegraphics{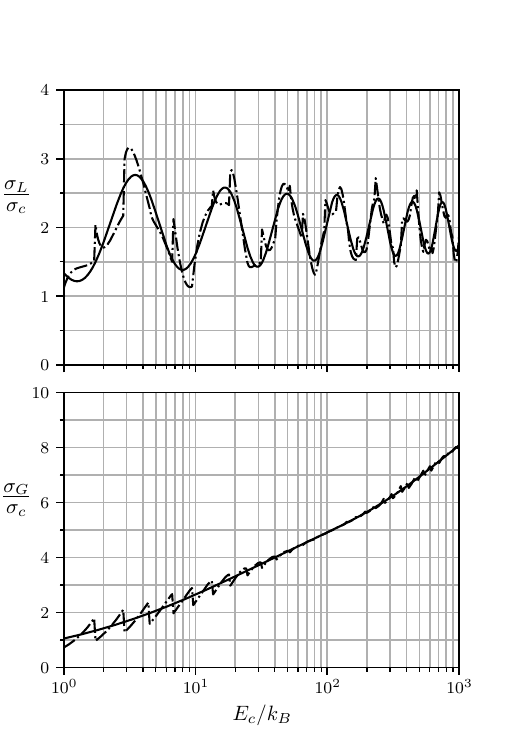}
\caption{\label{fig:sigmaQ} Contributions to the total scattering cross-section as a function of energy calculated via the integral given in the text (solid) or the summation (dashed line).  In both cases the $\eta_l$ is calculated by the WKB approximation.  The upper figure shows contributions to `Langevin' ($p<1$) collisions and lower plot `Glancing' ($p>1$) collisions.  In both cases the result is scaled by the classical value $\sigma_c=\pi b_c^2$.  The hard-sphere center was chosen to have a radius satisfying $\phi_R=0$.}
\end{center}
\end{figure}

The total scattering cross-section is given by \cite{bransden2003physics}
\begin{align}
\sigma_\mathrm{tot}&=\frac{4\pi}{k^2}\sum_{l=0}^\infty (2l+1)\sin^2(\eta_l),\\
&=(\pi b_c^2) \sum_{l=0}^\infty 8p \sin^2(\eta_l)\,\Delta p,\\
&\approx (\pi b_c^2) \int_0^\infty 8p \sin^2(\eta_l)\,dp
\end{align}
where we have used $\Delta p=p_{l+1}-p_l=1/(k b_c)$. This can be separated into contributions from $p<1$ ($l\leq l_0$) and $p>1$ ($l > l_0$), which we denote $\sigma_L$ and $\sigma_G$ respectively, in analogy with classical Langevin and glancing collisions.  These contributions are shown in Fig.~\ref{fig:sigmaQ} as calculated using either the integral or the discrete sum with $\eta_l$ determined by the WKB approximation in both cases.  Interestingly, $\sigma_L$ shows significant oscillation about a value that is two times the classical value.  The factor of two also appears in the quantum vs classical scattering by a hard sphere.  However, in this case, $R_0$ only changes the phase of the oscillation seen in Fig.~\ref{fig:sigmaQ}.  This is not so surprising given that $\eta_l$ only depends on the radius of the core through the energy-independent term $L/R_0$.  The dashed lines in Fig.~\ref{fig:sigmaQ} use the discrete sum, which shows prominent discontinuities in $\sigma_L$ when the number of angular momenta included in the sum changes with collision energy.

The approach in \cite{hankin2019systematic} used a randomized $\eta_l$ to model scattering properties in the Langevin region, which we will henceforth refer to as the randomized phase approximation (RPA).  Within this approximation, $\sin^2\eta_l$ may be replaced by 1/2 which leads to the same factor of two enhancement of the Langevin collision rate.  The approximation arises from the fact that $\eta_l$ is typically very large such that it appears random when taken modulo $\pi$, which would likely be true in general.  It is of interest then to explore this more carefully to understand the repercussions it has.

The Langevin regime applies to all angular momenta below a critical value $l_0\approx k b_c$ with a contribution to the scattering amplitude given by
\begin{align*}
f(\theta)=&\frac{1}{k}\sum_{l=0}^{l_0}(2l+1)e^{i\eta_l}\sin\eta_l\, P_{l}(\cos\theta),\\
=&\frac{1}{2 i k}\sum_{l=0}^{l_0}(2l+1)e^{2i\eta_l}P_{l}(\cos\theta)\\
&\qquad-\frac{1}{2 i k}\sum_{l=0}^{l_0}(2l+1)P_{l}(\cos\theta),
\end{align*}
where we have separated off a component that is independent of the scattering potential.  By virtue of the completeness property of Legendre polynomials, this term tends to a delta function with increasing $l_0$.  Thus it is predominantly a forward scattering component.  The first term has a relatively flat distribution for both the real and imaginary parts.  The size of the repulsive core only provides a fixed offset in the phase and hence does not change the magnitude.  However, the imaginary part will have some influence on the forward scattering component as the latter is pure imaginary.  On their own, either component gives a contribution to the total cross-section of $\sigma_c=\pi b_c^2$ and it is the interference between them that gives the oscillation in  $\sigma_L$.
    
The forward scattering component can be adequately approximated using 
\begin{equation}
\label{Eq:LegendreApprox}
P_l(\cos\theta)\approx \sqrt{\frac{\theta}{\sin\theta}}\, J_0\left(\left(l+\tfrac{1}{2}\right)\theta\right),
\end{equation}
which leads to
\begin{equation}
f(\theta)=i b_c\frac{J_{1}\left[k b_c\theta\right]}{\sqrt{\theta\sin\theta}}\approx i b_c\frac{J_{1}\left[k b_c\theta\right]}{\theta}.
\end{equation}
Most of this contribution is limited to scattering angles below the first zero of the Bessel function i.e. $\theta\lesssim 4/(l_0+\tfrac{1}{2})$.

The approximation given in Eq.~\ref{Eq:LegendreApprox} applied to glancing collisions with $l>l_0$ gives the expected result that this contribution is primarily forward scattered and limited to the same small $\theta$.  Although one cannot cleanly separate contributions to a differential cross-section, we can still identify three physical components: forward scattered Langevin at the classical rate $\sim \Gamma_L$, which primarily introduces a phase in the Ramsey signal; classical Langevin at the same rate $\sim\Gamma_L$, which will be suppressed due to the momentum transfer; and glancing collisions that occur at an enhanced rate but have minimal effect as they have neither a large contribution to the SODS or a clock state phase shift.

The momentum transfer cross-section $\sigma_M$ can be also be determined in the same way as above giving
\begin{align}
\sigma_M(k)&= \frac{4\pi}{k^2}\sum_{l=0}^\infty (l+1)\sin^2(\eta_{l+1}-\eta_{l}),\\
&\approx \pi b_c^2 \int_0^\infty 4 p\sin^2\left(\frac{1}{kb_c}\frac{d\eta}{dp}\right)dp
\end{align}
Because of the derivative, the constant term from the size of the sphere is eliminated so there is no variation with the size of the sphere.  Since non-constant terms in $\eta_l$ are proportional to $kb_c$, the dependence on $kb_c$ drops out except for the angular momentum states contributing to $p<1$.  Thus, there is no significant variation with energy.  The constant term in $\eta_l$ is the predominant difference between our result with that of \cite{dalgarno1958mobilities} and consequently we find a similar value of $\sigma_M\approx 1.1 \pi b_c^2$ despite differences in the WKB approximation.  That $\sigma_M$ is practically identical to the classical value $\pi b_c^2$ is consistent with the fact that half of the Langevin collisions and the majority of glancing collisions are forward scattered and, hence, do not contribute to momentum transfer.

\subsection{Quantum formulation of the clock shift}
In \cite{vutha2017collisional}, the authors derive the expression
\begin{equation}
\label{eq:QuantumShift}
\delta \omega_c=-n v_n\Im \int \bar{f}^*(\theta) f(\theta)d\Omega,
\end{equation}
for the clock shift where $\bar{f}(\theta)$ and $f(\theta)$ are, respectively, the excited and ground state scattering amplitudes.  In the absence of knowledge of the scattering potentials, the authors of \cite{vutha2017collisional} argue that the Cauchy-Schwarz inequality can be used to provide a conservative bound for the clock shift as the geometric mean of the total scattering rates.  On the basis of the given expression, this is a mathematical truth, but it is a considerable over-estimation given that the dominant contribution to the total scattering rate arises from collisions that do not penetrate the angular momentum barrier, and hence cannot significantly contribute to a shift.  Moreover, anything not forward scattered is heavily suppressed through non-closure of the Ramsey interrogation, which their derivation does not appear to consider.

In general, the scattering amplitude can be separated into three components 
\[
f(\theta)=f_0(\theta)+i f_\mathrm{fs}(\theta)+ f_G(\theta),
\]
where
\begin{align*}
f_0(\theta)&=\frac{1}{2ik}\sum_{l=0}^{l_0}(2l+1)e^{2i\eta_l}P_{l}(\cos\theta),\\
f_\mathrm{fs}(\theta)&=\frac{1}{2k}\sum_{l=0}^{l_0}(2l+1)P_{l}(\cos\theta),\\
\intertext{and}
f_G(\theta)&=\frac{1}{k}\sum_{l>l_0}(2l+1)e^{i\eta_l}\sin(\eta_l)P_l(\cos\theta).
\end{align*}
In the context of the short range hard-sphere model we have considered, $l_0$ is a critical value of $l$ separating glancing collisions, which do not penetrate the angular momentum barrier, from Langevin collisions that do. Moreover $f_G$ and $f_\mathrm{fs}$ would be the same for both states as $f_G$ only depends on the common $r^{-4}$ potential and $f_\mathrm{fs}$ has no dependence on the interaction potentials.  Real molecular potentials may include long range terms, such as $r^{-6}$, that contribute to the angular momentum barrier such that $l_0$, as a critical value, may not be the same for both states.  Nevertheless, the $r^{-6}$ would be typically dwarfed by the $r^{-4}$ term such that a value of $l_0$ common to both states can be found for which $f_G$ is approximately the same for both states.  Moreover, $l_0$ would, in general, be close to the critical value such that $f_G$ would capture the dominant contributions from glancing collisions.  With this separation, the clock shift becomes
\begin{multline}
\label{eq:QuantumShift0}
\delta \omega_c=-n v_n\Im \int \Big[\bar{f}_0^*(\theta) f_0(\theta)\\
+(\bar{f}_0^*-f_0^*)(f_G+i f_\mathrm{fs})\Big]d\Omega.
\end{multline}
If integration is taken over the full range of $\theta$, orthogonality of the Legendre polynomials eliminates the contribution from $f_G$.  Applying the Cauchy-Schwarz inequality to the remaining term leads to a bound of $\pm \Gamma_L$, which is just the WCCS.  Furthermore, it is generally true that the second term tends to vanish when averaged over a thermal distribution of collision energies even when the $\theta$ integration is restricted.  This can be illustrated using a hard-sphere model.

Within the WKB approximation, a hard-sphere repulsion defined by a phase $\phi_R$ has a scattering amplitude 
\[
f(\theta)=e^{2i \phi_R}f_0(\theta)+i f_\mathrm{fs}(\theta)+ f_G(\theta),
\]
where $f_0(\theta)$ is evaluated for a reference radius giving $\phi_R=0$.  With $\bar{\phi}_R$ and $\phi_R$ being the angles for the excited and ground states respectively, the clock shift becomes
\begin{align}
\label{eq:QuantumShiftHS}
\delta \omega_c=&-n v_n\Im \int \Big[e^{-2i(\bar{\phi}_R-\phi_R)}|f_0|^2\nonumber\\
&\quad+e^{-2 i\bar{\phi}_R}f_0^*(f_G+i f_\mathrm{fs})-e^{-2 i \phi_R}f_0^* (f_G+i f_\mathrm{fs})\Big]d\Omega,\nonumber\\
=&-n v_n\Im\Big[e^{-2i(\bar{\phi}_R-\phi_R)}A_0\nonumber\\
&\qquad+(e^{-2 i\bar{\phi}_R}-e^{-2 i\phi_R})A_\mathrm{+}\Big].
\end{align}
This expression should be averaged over the velocity distribution of the incoming neutral and integrations restricted to angles that do not give sufficient momentum to the ion to affect the final Ramsey pulse.  For the latter, it is sufficient to approximate the RSF shown in Sect.~\ref{sectI} as a step function with a cutoff at $\bar{v}_c$, limiting the angles to 
\[
\theta_c=\begin{cases} 2\sin^{-1}\left(\frac{\bar{v}_c\sqrt{2}}{\bar{v}_n}\right), & \bar{v}_n>\bar{v}_c\sqrt{2},\\
\pi, & \bar{v}_n<\bar{v}_c\sqrt{2}\end{cases},
\] 
which is typically small for room temperature velocities of the neutral.  

\begin{figure}[t]
\begin{center}
\includegraphics[width=0.95\linewidth]{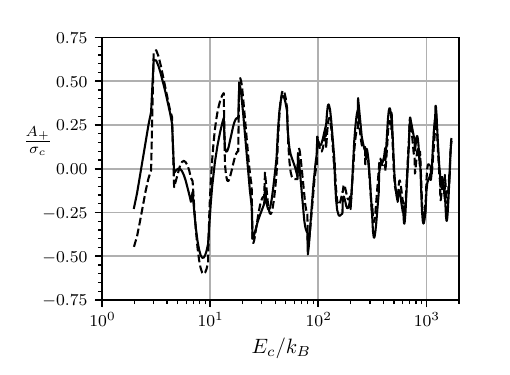}
\caption{\label{fig:ApHS} Contribution to the clock shift from $\Re(A_+)$ as a function of collision energy.  A similar behaviour occurs for $\Im(A_+)$.  The dashed curve includes all angles $\theta$, whereas the solid curve includes a cutoff to account for the RSF as noted in the main text.}
\end{center}
\end{figure}

As a function of collision energy, the real and imaginary parts of $A_+$ oscillate about zero in a similar manner with an amplitude of $\sim 0.3\sigma_c$ and this does not significantly depend on $\theta_c$.  This is illustrated in Fig.~\ref{fig:ApHS}, which shows the variation of $\Re(A_+)$ with collision energy when restricting the angles of integration (solid) or not (dashed).  This is not surprising given that contributions to $A_+$ are primarily in the forward direction.  Moreover, when averaging over a thermal distribution, both the real and imaginary parts of $A_+$ are heavily suppressed giving a magnitude of $\sim 5\times 10^{-4}\sigma_c$ for integration over collision energies in the range $k_B\times (2,1700)\,\mathrm{K}$, which includes 99\% of the MB distribution at 300\,K.  Hence $A_+$ can be neglected leaving only $A_0$, which, in contrast, is more uniformly spread over all angles $\theta$, does not significantly oscillate with the collision energy, and the relevant collision rate is the classical value $\Gamma_L$.

For comparison with \cite{vutha2017collisional,davis2019improved} we can integrate over all scattering angles and take the largest and smallest possible values allowed by $\bar{\phi}$, and $\phi$ as a bound on the clock shift.  Noting that $A_+$ does not contribute and $A_0\approx \sigma_c$, averaging over the thermal distribution and taking $\bar{\phi}-\phi = \pm \pi/4$ gives the expected clock shift bound of $\pm\Gamma_L$.  As noted earlier, this is within $\sqrt{2}$ of the clock shift uncertainty as that given in \cite{davis2019improved} for Al$^+$, which was derived using calculated molecular potential curves for the AlH$_2^+$ complex, and also the WCCS given by Eq.~\ref{WCCS} assuming the classical Langevin collision rate.

The bound given in \cite{davis2019improved} is overly pessimistic if the integration is not restricted in the range of angles in accordance with the fact that most collisions at large angles decouple the ion from the clock laser removing the interrogation from the Ramsey signal.  As noted, $A_+$ averages to near zero with or without a cutoff in $\theta$.  Since contributions to $A_0$ are distributed somewhat evenly over $\theta$ this substantially reduces the bound.  For $\bar{v}_c=0.04$, the bound is reduced to $0.04\Gamma_L$.  This is very much consistent with Eq.~\ref{eq:RSCSapproxB} and the classical argument in the previous section, which showed that the recoil velocity distribution is approximately a constant and on the order of unity for small $\bar{v}$.
\subsection{Lennard-Jones Potentials}
That the result given in \cite{davis2019improved} is the same as a naive estimate taking a classical Langevin collision rate $\Gamma_L$ and assuming the worse case shift of the Ramsey fringe raises the question of how well one can realistically do in simulating molecular collision potentials to accurately assess a possible collision shift.  We have investigated this by taking a Lennard-Jones (LJ) potential in addition to the $r^{-4}$ potential as defined in Appendix~\ref{LJPotential}.  The potential is parameterized such that it has a minimum of $V_0=-(1+x) C_4/(6 r_m^4)$ at $r=r_m$ with a $C_6$ coefficient given by $C_6=(x-1)r_m^2 C_4/3$.  For an ion-neutral collision we expect the $C_6$ coefficient to be largely independent of the clock state \cite{cote2000ultracold}.  

\begin{figure}[t]
\begin{center}
\includegraphics{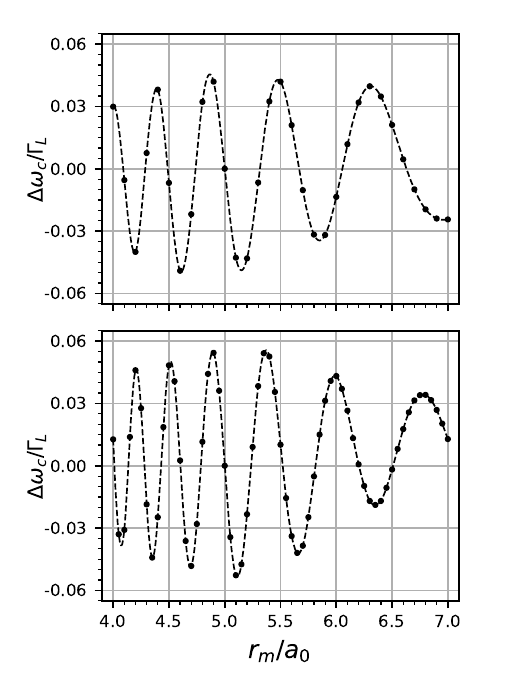}
\caption{\label{fig:LJShifts} Variations in the clock shift calculated from Lennard-Jones potentials defined in Appendix~\ref{LJPotential} using a cutoff of $\bar{v}_c=0.04$.  One clock state has $r_m=5 a_0$  and $x$ chosen to fix the value $C_6$.  For the second clock state $r_m$ and $x$ lie on the same contour of fixed $C_6$.  Upper plot uses $C_6\approx 0.83 C_4$ in atomic units ($x=1.1$ for the fixed clock state).  Lower plot uses $C_6\approx 8.33 C_4$ in atomic units ($x=2$ for the fixed clock state).  Dots are the calculated shifts and the dashed curves are a cubic spline to guide the eye.}
\end{center}
\end{figure}

Collision shifts for a range of potentials are plotted in Fig.~\ref{fig:LJShifts}.  With the expectation that the minimum of either scattering potential would be in the range of $4 a_0\sim7 a_0$, we calculate the collision shift with one clock state parameterised by $r_m=5a_0$ and $x$ chosen to fix the $C_6$ coefficient relative to $C_4$.  For the second clock state we allow $r_m$ to vary between $4a_0$ and $7a_0$ with $x$ chosen to keep $C_6$ constant i.e. it lies on the same contour of $C_6$ as the first clock state.  In both cases, $C_4$ is taken to be the rotationally averaged value.  The clock shift is estimated taking $\bar{v}_c=0.04$ to determine the cutoff $\theta_c$ when averaging over the thermal distribution.  

For molecular hydrogen, we expect $C_4\approx C_6$ when expressed in atomic units \cite{miliordos2018dependence}.  Consequently, the upper plot uses $C_6 \approx 0.833 C_4$ or $x=1.1$ for the clock state having a fixed potential.  As we expect a relative insensitivity to the value of $C_6$, the lower plot uses $x=2$ for the clock state having a fixed potential, which makes $C_6$ a factor of 10 larger.  When the potentials are identical, the clock shift will be zero.  Thus changes in the clock shift as one potential is changed is an indicator of the sensitivity the shift has on the accuracy at which the molecular potentials can be specified.

Plots of the calculated clock shifts for the two contours considered are shown in Fig.~\ref{fig:LJShifts}.  Calculation of $\eta_l$ values were carried out numerically up to values of $l$ for which $\eta_l<0.01$.  At this point $\eta_l$ is well approximated by Eq.~\ref{WKBbelow} although inclusion of additional terms in the summation for $f(\theta)$ makes little difference to the plots.  Truncated integrals over $\theta$ made use of the analytic approximations given in appendix~\ref{sect:TLI}.

As seen in Fig.~\ref{fig:LJShifts}, variations of the potentials lead to maximum variations of the collision shift that are consistent with those determined by the hard-sphere model and with Eq.\ref{eq:RSCSapproxB} using the classical recoil velocity distribution derived in Sect.~\ref{sect:Classical} i.e. $\Delta \omega_c\approx \Gamma_L \bar{v}_c$.  The sinusoidal-like variations along the contour can be understood in the context of Eq.~\ref{eq:QuantumShiftHS} and the WKB approximation used to calculate $\eta_l$.  For small variations of the potentials, the lowest order variation of $\eta_l$ will be a simple constant, which will change the contributions to the imaginary part of the clock shift integral.  When the potential is relatively deep, the change in $\eta_l$ for those angular momentum penetrating the barrier will be substantial for even modest changes in the potential leading to more rapid oscillations in the clock shift as clearly seen for smaller $r_m$ in the plots.  This behaviour does not depend on the use of a cutoff in the $\theta$ integral and explains why the bounds in \cite{davis2019improved,leibin2025collisional} are consistent with $\pm \Gamma_L$ when not restricting $\theta$.  Different modelling strategies lead to significant changes in the phase of the clock shift integral. 

There is some variation in the maximum and minimum values in Fig~\ref{fig:LJShifts}.  This is expected as the exact recoil velocity distribution depends on the underlying differential cross-section which will change as the potentials are varied.  The phase shifts $\eta_l$ for the Lennard-Jones potential and the hard-sphere model have qualitatively the same form.  We would expect this to be the case for any potential in which the long range behaviour is governed by an attractive $r^{-4}$ potential based on the mathematical form of the WKB integrals.  For the same reason, we would expect the underlying recoil velocity distribution to have the same qualitative features as shown in Sect.~\ref{sect:Classical},  specifically that the PDF in the neighbourhood of zero is $\kappa\approx 1$.  If $\kappa$ were substantially larger, it would require $\tilde{v}$ to change significantly, which is not compatible with it being determined by conservation of energy and momentum in the collision.  This is consistent with the intuitive expectation that details of the intermolecular potentials should be largely irrelevant for collisions at the thermal energies of interest.

As evident from the plots, the maximum and minimum values do not significantly change when the $C_6$ coefficient is increased even by an order of magnitude.  This is not surprising given that the $C_6$ term drops off more rapidly suppressing its contribution relative to the $C_4$.  More significant shifts can be seen if $C_6$ has a substantial difference between the clock states.  As noted, we would not expect this to be the case for ion clocks.  A dependence on the clock state would come from a Van der Waal interaction, with the dependence on the clock state coming from differences in the clock-state polarizabilities.  Even if this is significant, which it is not for Lu$^+$, it would be dwarfed by the $C_4$ coefficient within the WKB integral.
\subsection{Real diatomic molecules}
As in \cite{hankin2019systematic}, we have treated the collision shift considering only the scalar interaction of the ion and a polarizable neutral neglecting any rotational motion.  It would be remiss to not mention the possible influence of the tensor polarizability, permanent quadrupole, and rotational energy inherent to diatomic molecules.  The tensor polarizability and permanent quadrupole introduce a non-central component to the collision potential with the later having an $r^{-3}$ scaling.  The rotational energy introduces the possibility of rotational coupling.  The quadrupole term, in particular, has been shown to signifcantly alter capture rates in cold collisions between ions and either H$_2$ or N$_2$ in a manner that depends on the sign of the quadrupole moment \cite{zhelyazkova2022multipole}.  

In atomic units, the classical long range potential between an ion and a diatomic molecule is given by
\begin{equation}
V(r)=-\frac{\alpha_m}{2 r^4}+\left(-\frac{\Delta\alpha}{3 r^4}+\frac{Q_{zz}}{2 r^3}\right)P_2(\cos\theta)
\end{equation}
where $\theta$ is the angle between internuclear axis of the molecule and the electric field of the ion (the collision axis), $Q_{zz}\approx 0.95\,\mathrm{a.u.}$ is the quadrupole moment along the molecular axis, and $\alpha_m\approx 5.4\,\mathrm{a.u.}$ and $\Delta\alpha\approx 2.1\,\mathrm{a.u.}$ are determined from the polarizability tensor (see appendix~\ref{tensors} for definitions).

In the case of H$_2$, rotational states $\ket{J,M}$ have a large separation such that only the lowest energy rotational state is occupied at 300\,K: $J=0$ for para-hydrogen (p-H$_2$), which constitutes 25\% of H$_2$ at 300\,K, or $J=1$ for ortho-hydrogen (o-H$_2$), which is the remaining 75\%.  Consequently the quantum mechanical nature of rotations cannot be ignored.  The most straightforward simplification is to assume the molecule adiabatically follows the collision axis and treat the tensor component as a perturbation to a rigid rotor.  Due to the large splittings in the rotational energies it is sufficient to take the lowest order perturbation, which eliminates the tensor term for p-H$_2$ and gives
\begin{equation}
V(r)=-\frac{\alpha_m}{2 r^4}+\left(-\frac{\Delta\alpha}{15 r^4}+\frac{Q_{zz}}{10 r^3}\right)(2-3M^2)
\end{equation}
for o-H$_2$.  The polarizability components of this equation are consistent with calculations of the scalar and tensor polarizabilities in \cite{sun2021polarizabilities}, which do not use the Born-Oppenheimer approximation.  

The quadrupole term is most significant for the $M=0$ state of o-H$_2$ and gives rise to a slight repulsion with an energy barrier less than 2\,K, which is responsible for the decrease in capture rates in cold collisions \cite{zhelyazkova2022multipole}.  For the thermal energies considered here, this is unlikely to be significant and if anything would tend to suppress the rate determining the clock shift bound. The opposite is true for the $M=\pm1$ states but the effect is reduced by a factor of two.  Thus any influence of the tensor component on the rate determining the clock shift bound would tend to cancel when statistically averaged over the populations.  This is supported by the approachs reported in \cite{davis2019improved,leibin2025collisional}, which account for orientations of the hydrogen molecule in different ways.  Thus the reasonable consistency of the bounds given in each case with the WCCS would suggest the tensor term can be ignored, at least to the accuracy one can estimate the collision shift from the potentials.  A more detailed consideration of the effect of the tensor term is beyond the intended scope of this work and will be the subject of future investigation.

One may well ask if the adiabatic approximation used for the tensor term is valid.  The time scale associated with rotations would be roughly $\tau\sim I/\hbar$ where $I$ is the moment of inertia of H$_2$.  The maximum rate of change of the collision axis in a space fixed frame would be at the minimum approach for those collision near the critical impact parameter.  The angle change in time $\tau$ would then be $v_n\tau\sqrt{2}/b_c\sim 0.25$, where we have taken $v_n=\tilde{v}_n$ and $b=b_c\approx 10a_0$.  Thus non-adiabaticity may well be important in any attempt to accurate calculate a collision shift.  We emphasise that a bound on the collision shift is determined by the collision rate, which may be accurately estimated from calculations, but the collision shift itself depends on phase of the integral in Eq.~\ref{eq:QuantumShift} and that places much more stringent demands on the accuracy of calculated collision dynamics.

\section{Discussion}
Given the consistency obtained with a purely classical analysis of Langevin scattering and a quantum treatment incorporating either a short-range hard-sphere interaction or a Lennard-Jones type potential, a reasonable bound for the collision shift of the clock transition would be 
\begin{equation}
\frac{\delta f_c}{f_c}=\frac{\kappa \Gamma_L \bar{v}_c}{2\pi f_c},
\end{equation}
where $\kappa\approx 1$, $\Gamma_L$ is the classical Langevin collision rate, and $\bar{v}_c$ is determined from the angular average of Eq.~\ref{eq:carrier} appropriately modified for the geometry of the ion-trap and parameters of the probe for the system under consideration.  This expression captures the simple, intuitive expectation that a collision shift can be bounded by a classical capture rate for such collisions reduced by the fact that most collisions effectively remove the interrogation from the spectroscopy signal at least on average.

The parameter $\kappa$ will depend on the recoil velocity distribution.  From our classical analysis the relevant velocity distribution is roughly a half gaussian distribution $e^{-\bar{v}^2}$ giving $\kappa=2/\sqrt{\pi}\approx 1.12$.  Thus $\kappa\approx1$ would be sufficient for most purposes.  The cut-off $\bar{v}_c$ is a mathematical property of $\mathcal{R}$ and, for two qualitatively different velocity distributions, it was shown to be consistent to take $\bar{v}_c\approx 0.034$ for the geometry we used. This is close to the point at which $\mathcal{R}=0.5$.  We therefore use this 50\% threshold condition to define $\bar{v}_c$.  This leads to a collision shift bound of
\[
\frac{\delta f_c}{f_c}\approx \pm 6.2 \times 10^{-21} (\mathrm{nPa})^{-1}
\]
for $^{176}$Lu$^+$ in a 300\,K background gas of molecular hydrogen.  From consideration of Lennard-Jones potentials, improving on this bound would require molecular potentials to be calculated to at least a few percent accuracy, which would likely need to include the quantum mechanical nature of molecular rotations, any dependence of the potentials on collision energy, and possibly even the dynamical nature of the collision.  

What remains is a determination of the vacuum pressure or equivalently the collision rate $\Gamma_L$.  For all practical purposes the gauge attached to chamber and accounting for pumping speed restrictions to the ion ought to suffice at least for the measurement precision that can be obtained in any foreseeable future.  However, the fact that the atom-laser coupling is the most significant effect affords a possible alternative, at least for clock systems that are not lifetime limited.  One can simply shelve the ion into the dark state, wait and re-shelve to the bright state.  Failure to reshelve, even after multiple attempts, would register a collision event.  There will be a small number of Langevin collision events that minimally influence the atom-laser coupling, but this would be a rather small underestimate in determination of the rate.  More importantly, it would be overcompensated by glancing collisions near to the critical impact parameter that would decouple the ion from the clock laser even though they minimally impact on the collision shift.  

The overestimate in the collision rate from glancing collisions can be estimated from the classical description, noting that a classical description should be valid for scattering angles required to impart significant momentum to the ion. Using Eq.~\ref{eq:classical_cutoff} with a cutoff of $\delta_c=\bar{v}_c^2$, the rate could be overestimated by as much as a factor of three for the case of $^{176}$Lu$^+$ considered here.  We also note that there would be a number of collisions that not only affect the clock laser coupling, but also influence detection.  We have not considered this here as the clock laser is significantly decoupled from the ion at recoil energies well below those at which detection is affected.  However, the influence on detection may provide additional indicators of the behaviour after collisions and will be the subject of future investigation. 

In \cite{hankin2019systematic}, the authors provide a frame work in which to assess the collisional shift for ion-based clocks.  In many respects, our treatment captures the same essential physics, but application to a single ion allowed analytic expressions to be derived.  Beyond that, the most significant differences in the formulations are: the treatment of motional states arising from a collison, Rabi vs Ramsey spectroscopy, and the application to an ion crystal vs a single ion. It is reasonable to ask to what degree these differences are important.  

In \cite{hankin2019systematic}, the internal dynamics subsequent to a collision uses projections of the motional state onto Fock states.  For a Fock state, coupling to the laser will be reduced by $e^{-\eta^2/2}L_n^0(\eta^2)$, where $L_n^0(\eta^2)$ is the associated Laguerre polynomial. To a very good approximation \href{https://dlmf.nist.gov/18.15}{DLMF\S18.15},
\[
e^{-\eta^2/2}L_n^0(\eta^2)\approx J_0\left(2\eta\sqrt{n+\tfrac{1}{2}}\right), 
\]
which connects to the modulated laser interpretation.  For a coherent state $\ket{\alpha}$, the mean and variance of the number operator is $|\alpha|^2$.  As $\alpha$ is almost always large, $n+1/2$ can be replaced by $|\alpha|^2$ giving the correct result consistent with a laser modulation.  What is less clear is how subsequent evolution in the crystal is handled.

In an ion crystal, typical collisions from a background gas disrupt the crystal and that very fact is used in \cite{hankin2019systematic} and by others \cite{pagano2018cryogenic,hausser2025in+} to estimate collision rates and background gas pressures.  Under such conditions, the motion will be highly non-linear, well outside the harmonic regime from which the Fock states are taken, and may change a crystal configuration in a non-trivial way.  Moreover, the energy imparted to the ion is dispersed to the crystal as a whole on a timescale much faster than typical interrogation times.  Thus an impact that is sufficient to reconfigure the crystal and decouple the ion from the laser, can result in the ion being locked into a new configuration and at least partially re-coupled to the laser, in effect reducing the modulation index arising from the motion.  What Fock states mean in this scenario is not clear.  

A better approach for crystals would be a semiclassical one in which the motion is propagated by the classical laws of motion, as was done in \cite{hankin2019systematic}, but that motion should be mapped to a time-dependent detuning, or equivalently a phase $e^{ikx}$, in the atom-laser coupling.   For a single ion confined in a harmonic potential, that approximation is exact and equivalent to what we have done.  For ion crystals, it would eliminate sampling of Fock states and dramatically reduce the amount of trials needed for a Monte-Carlo simulation.  

The degree to which our results can be mapped to systems using multiple ions is beyond the scope of this work.  Factors that would need to be carefully investigated are the increased collision rate with the number of ions, the decrease in modulation index associated with dispersion of the recoil energy across the crystal, and the vibrational mode structure.  A fair guess for the system reported in \cite{hankin2019systematic}, would be to treat the recoil of the crystal as a whole and assume the resulting recoil energy is evenly divided between the two ions to determine the recoil velocity appropriate to the logic ion.  With $\lambda=269\,\mathrm{nm}$, and $\omega_x=\omega_y=2\pi\times 3.5\,\mathrm{MHz}$, this leads to $\bar{v}_c\approx 0.029$ for the 50\% threshold for $\mathcal{R}$.  The collision rate would be proportional to the number of ions and the pressure, which doubles the collision rate to account for the two ions and 57\,nPa bounds the specified background pressure of 38(19)\,nPa.  This leads a bound of $\pm 1.7\times 10^{-19}$, which is in fair agreement of their value of $\pm 2.4\times 10^{-19}$ given the crudeness of the estimate.

We have considered Ramsey verses Rabi interrogation.  However, we would not expect that to change results significantly.  Inasmuch as the SODS subsequent to a collision can be neglected and the effect on the laser coupling roughly approximated as a step function of the recoil velocity, a similar line of reasoning used here can be applied.  For sufficiently large momentum transfer, the ion is effectively decoupled from the laser, which is equivalent to the pulse time being randomly changed and there is no associated shift.  As for Ramsey spectroscopy, this interpretation is to be taken in the context of averaging over recoil orientations.  For the small portion of collisions in which the laser remains coupled, the effect is only to switch the phase by $\pi/2$ within the pulse and the effect on the spectroscopy signal can be calculated analytically \cite{rosenband2008frequency}.  Glancing collisions have no effect beyond a negligible reduction in contrast, so collision rates would be still based on the classical Langevin rate.  This is consistent with the formulations in \cite{vutha2017collisional,davis2019improved}, which do not differentiate between interrogation techniques.

Finally, it is worthwhile to consider the collision shift in the context of a physical realization of a Monte-Carlo simulation, which is the actual operation of a clock.  Every interrogation is a trial of a Monte-Carlo simulation.  A Lu$^+$ clock comparison operating with a 90\% contrast and a 5\,s interrogation time will integrate down to mid $10^{-19}$ in 200\,hours of operation.  In that time there would be approximately 144,000 Monte-Carlo trials.  With a collision rate of 0.001/s, $\sim 720$ of these interrogations would incur a collision and $\approx95\%$ of these would be effectively removed from the Ramsey signal via $\mathcal{R}$, leaving $\sim 36$ interrogations to influence the final result.  There is simply not enough trials for a problem to manifest.
\acknowledgements
The authors wish to thank Berge Englert for valuable insights and discussions.  This project is supported by the National Research Foundation, Singapore through the National Quantum Office, hosted in A*STAR, under its National Quantum Engineering Programme 3.0 Funding Initiative (W25Q3D0007) and under its Centre for Quantum Technologies Funding Initiative (S24Q2d0009).
\appendix
\section{Potential curve models}
\label{LJPotential}
For a model of the molecule potential, we can take a Lennard-Jones potential in addition to the attractive $r^{-4}$ potential.  Taking
\[
V(r)=A\left(\frac{r_m}{r}\right)^{12}-B\left(\frac{r_m}{r}\right)^6-\frac{C_4}{2 r^4},
\]
with
\[
A=V_0-\frac{C_4}{6r^4_m},\quad B=2\left(V_0-\frac{C_4}{3r^4_m}\right)
\]
gives a potential with a minimum $-V_0$ at $r=r_m$.  We constrain the choice of $V_0$ and $r_m$ so that $A,B>0$.  To this end, it is convenient to set $V_s=C_4/(6 r_m^4)$ and $V_0=(x+1)V_s$.  The potential can then be written
\[
V(r)=\left[x \left(\frac{r_m}{r}\right)^{12}-2(x-1) \left(\frac{r_m}{r}\right)^6-3 \left(\frac{r_m}{r}\right)^4\right]V_s.
\]

Near to the origin, only the $r^{-12}$ term is important so the solution to the radial equation is asymptotically
\[
u(r)\sim\sqrt{\bar{r}}\,K_{1/10}\left(\frac{1}{5 (\bar{r})^5}\right)
\]
where $K_n(z)$ is the modified Bessel function of the second kind and
\[
\bar{r}=\left(\frac{3 r_m^2}{x L^2}\right)^{1/10}\left(\frac{r}{r_m}\right).
\]
This approximate solution is practically zero at any point inside the boundary.  It is then sufficient to integrate the radial equation with $u(r_0)=0$ and $u'(r_0)\neq 0$ where $r_0$ is any point less than the classical turning point.  The phase shift $\eta_l$ can then be determined from the asymptotic form $\sin(k r-l\tfrac{\pi}{2}+\eta_l)$.  If the classical turning point is at $r_T$, a suitable choice of $r_0$ is about 85-90\% of $r_T$.  Less than this and the integration can become ill-behaved, and values near $r_T$ can alter the value of $\eta_l$.  Note that convergence to the asymptotic form is rather slow but it is sufficient to check that estimates from the WKB approximation are reasonable.  More precise calculations might be possible using the approach given in \cite{sesma2013exact}, which is similar in spirit to the construction in \cite{idziaszek2011multichannel} for the $r^{-4}$ potential.
\section{Truncated Legendre Integrals}
\label{sect:TLI}
To facilitate numerical evaluation of Eq.~\ref{eq:QuantumShift} when restricting the range of integration over $\theta$, we make use of some analytic approximations based on Eq.~\ref{Eq:LegendreApprox} and symmetries of the Legendre polynomials.  We first define
\begin{equation}
\label{Eq:LegendreInt}
f_{l_1,l_2}(\theta)=\int_0^\theta P_{l_1}(\cos\theta')P_{l_2}(\cos\theta')\sin(\theta')d\theta'.
\end{equation}
which is symmetric in $l_1$ and $l_2$ so we may assume $l_1\geq l_2$.  For small $l_1,l_2$ the integral can be easily carried out analytically.  For other values, we make use of Eq.~\ref{Eq:LegendreApprox} and symmetry properties of Eq.~\ref{Eq:LegendreInt}.  Using Eq.~\ref{Eq:LegendreApprox} in Eq.~\ref{Eq:LegendreInt} gives the analytic result
\begin{widetext} 
\begin{align}
\bar{f}_{l_1,l_2}(\theta)&=\int_0^\theta \theta' J_0\left[\left(l_1+\tfrac{1}{2}\right)\theta'\right]J_0\left[\left(l_2+\tfrac{1}{2}\right)\theta'\right]d\theta'\nonumber\\
&=\begin{cases}
\frac{\left[\left(l_1+\tfrac{1}{2}\right)J_0\left[\left(l_2+\tfrac{1}{2}\right)\theta\right]J_1\left[\left(l_1+\tfrac{1}{2}\right)\theta\right]-\left(l_2+\tfrac{1}{2}\right)J_0\left[\left(l_1+\tfrac{1}{2}\right)\theta\right]J_1\left[\left(l_2+\tfrac{1}{2}\right)\theta\right]\right]}{\left(l_1+\tfrac{1}{2}\right)^2-\left(l_2+\tfrac{1}{2}\right)^2}\theta, & l_1\neq l_2\\
\tfrac{1}{2}\theta^2\left(J_0^2\left[\left(l_1+\tfrac{1}{2}\right)\theta\right]+J_1^2\left[\left(l_2+\tfrac{1}{2}\right)\theta\right]\right), & l_1 = l_2,
\end{cases}
\end{align}
which is a good approximation to Eq.~\ref{Eq:LegendreInt} only over a range of $\theta$ that depends on $l_k$, but is typically good for $l_1,l_2\geq 1$ and $\theta\leq \pi/4$.

For $l_2=0$, Eq.~\ref{Eq:LegendreInt} can be expressed in terms of Legendre polynomials, which may then be approximated via Eq.~\ref{Eq:LegendreApprox}.  This leads to
\begin{equation}
f_{l,0}(\theta)\approx \begin{cases}
\frac{1}{2l+1}\left(J_0\left[\left(l-\tfrac{1}{2}\right)\theta\right]-J_0\left[\left(l+\tfrac{3}{2}\right)\theta\right]\right)\sqrt{\frac{\theta}{\sin\theta}}, & \theta<\pi/2\\
(-1)^{l+1}f_{l,0}(\pi-\theta), & \theta>\pi/2,
\end{cases}
\end{equation}
which we use for $l>2$.  For $l_1\geq l_2\geq 1$ we use
\begin{equation}
f_{l_1,l_2}(\theta)\approx \begin{cases}
\bar{f}_{l_1,l_2}(\theta), & \theta\leq \pi/4,\\
\frac{1}{2l_1+1}\delta_{l_1,l_2}+\tfrac{1}{2}\left[\bar{f}_{l_1,l_2}(\theta)-(-1)^{l_1+l_2}\bar{f}_{l_1,l_2}(\pi-\theta)\right], &  \pi/4<\theta<3\pi/4,\\
\frac{2}{2l_1+1}\delta_{l_1,l_2}-(-1)^{l_1+l_2}\bar{f}_{l_1,l_2}(\pi-\theta), & \theta> \pi/4.
\end{cases}
\end{equation}
\end{widetext}
\section{The hydrogen polarizability}
\label{tensors}
Due to symmetry, the polarizability tensor for H$_2$ has two components $\alpha_{xx}=\alpha_{yy}=\alpha_\perp$ and $\alpha_{zz}=\alpha_\|$.  This may be expressed in the form $\alpha_m \mathbf{1}+\tfrac{2}{3}\Delta\alpha\mathbf{T}$ where $\alpha_m=(\alpha_\|+2\alpha_\perp)/3$, $\Delta \alpha=(\alpha_\|-\alpha_\perp)$, $\mathbf{1}$ is the identity and $\mathbf{T}$ is diagonal and traceless with non-zero components $\mathbf{T}_{xx}=\mathbf{T}_{yy}=-1/2$ and $\mathbf{T}_{zz}=1$.  This decomposes the polarizability into the irreducible scalar and tensor components.  The quadrupole operator is similarly diagonal and given by $\mathbf{Q}=Q_{zz}\mathbf{T}$.  The quadrupole operator is unfortunately subject to convention choices in which a factor of 1/2 is included \cite{miliordos2018dependence} or not \cite{torres2006torques}.  It is also common to speak of the quadrupole moment as a scalar without stipulating what that means. For the removal of doubt, we use the convention given in \cite{torres2006torques} with $Q_{zz}$ giving the quadrupole moment.  Values of the polarizability and quadrupole moment $Q_{zz}$ are taken from \cite{zhelyazkova2022multipole} accounting for convention choices in $\mathbf{Q}$.

Specification of polarizability can be given in atomic units, cgs units (cm$^3$), or SI units.  For the readers convenience, conversion factors from atomic units to cgs or SI are $a_0^3$ or $4\pi\epsilon_0 a_0^3$ respectively, where $a_0$ is the Bohr radius.  In addition, some authors report polarizabilities with a unit m$^3$/mol \cite{nie2025polarizability}. We assume this to be a molar refractivity with the leading order relation to polarizability given by $4\pi N_A \alpha_m a_0^3/3$, where $N_A$ is Avogadro's constant and $\alpha_m$ is the mean polarizability of the molecule (in atomic units).  

\bibliography{LuCollisions}
\end{document}